\documentclass{aa}
\bibpunct{(}{)}{;}{a}{}{,}
\usepackage[varg]{txfonts}
\usepackage{natbib,twoopt}
\usepackage{graphicx}   
\usepackage{multirow}
\usepackage{placeins}

\newcommand{\Msun}[0]{\rm{M_\odot}}
\newcommand{\FeH}[0]{\rm{[Fe/H]}}
\newcommand{\MH}[0]{\rm{[M/H]}}

\newcommand{\Teff}[0]{T_\mathrm{eff}}
\newcommand{\DtV}[0]{D_\mathrm{T,VSA19}}
\newcommand{\Dtfe}[0]{D_{T,{\rm Fe}}}

\begin{document}

\title{Impact of radiative accelerations on the stellar characterization of FGK-type stars using spectroscopic and seismic constraints}
\author{Nuno Moedas\inst{1,2}\thanks{nmoedas@astro.up.pt} \and Morgan Deal\inst{3} \and Diego Bossini\inst{4,5}}

\institute{ Instituto de Astrof\'isica e Ci\^encias do Espaço, Universidade do Porto, CAUP, Rua das Estrelas, PT4150-762 Porto, Portugal
\and Departamento de F\'isica e Astronomia, Faculdade de Ci\^encias da Universidade do Porto, Rua do Campo Alegre, s/n, PT4169-007 Porto, Portugal,  \and LUPM, Universit\'e de Montpellier, CNRS, Place Eug\`ene Bataillon, 34095 Montpellier, France \and Dipartimento di Fisica e Astronomia Galileo Galilei, Università di Padova, Vicolo dell’Osservatorio 3, I-35122 Padova, Italy \and Osservatorio Astronomico di Padova – INAF, Vicolo dell’Osservatorio 5, I-35122 Padova, Italy}

\date{Received XXX / Accepted YYY}

\abstract 
{Chemical transport mechanisms are fundamental processes in stellar evolution models. They are responsible for the chemical distribution, and their impact determines how accurately we can characterize stars. Radiative accelerations are one of these processes. They allow the accumulation of elements at different depths in the star.}
{We aim to assess the impact of radiative accelerations on the modeling of FGK-type stars and their impact on the prediction of surface abundances.}
{To reduce the cost of the computation of radiative accelerations, we implemented the single-valued parameters (SVP) method in the stellar evolution code MESA. The SVP method is more efficient in calculating radiative accelerations, which enables computations of large enough grids of models for stellar characterization.}
{Compared to models that include atomic diffusion (with only gravitational settling), the inclusion of radiative accelerations has a small effect on the inference of fundamental properties, with an impact of 2\%, 0.7\%, and 5\% for mass, radius, and age. However, the treatment of radiative accelerations is necessary to predict the chemical composition of and accurately characterize stars.}
{} 


\titlerunning{Impact of radiative accelerations on stellar characterization}
\authorrunning{Nuno Moedas et al.}
\maketitle

\section{Introduction}

\label{sec:Intro}

Chemical composition is a fundamental component in stars, and it undergoes significant variations through their evolution. Chemical transport mechanisms are mostly responsible for these changes. Therefore, knowing how to model these processes in stellar evolution codes is essential for the accurate characterization of stars. 
Spectroscopic missions such as the APO Galactic Evolution Experiment (APOGEE; \citealt{Ahumada2020}) and the Echelle SPectrograph for Rocky Exoplanet and Stable Spectroscopic Observations (ESPRESSO; \citealt{ESPRESSO}) have provided new and precise surface chemical abundances of stars, complemented by asteroseismic data from missions such as \textit{Kepler}/K2 \citep{kepler} and the Transiting Exoplanet Survey Satellite (TESS; \citealt{TESS}). These new constraints allow us to improve the models used for stellar characterization for upcoming missions such as PLAnetary Transits and Oscillations of Stars (PLATO; \citealt{Rauer2024}) and Ariel \citep{Tinetti2018,tinetti2021}.

There are a variety of chemical transport mechanisms, either microscopic or macroscopic, that compete to redistribute chemical elements, affecting the internal structure and evolution of stars. One of the most commonly included in stellar models is atomic diffusion. It is a microscopic process driven by chemical, thermal, and pressure gradients inside stars \citep{michaud15}, and it can be divided into two main components. One is gravitational settling, which brings the elements from the surface to the interior of the star, except for hydrogen, which is transported from the interior to the surface. The other are the radiative accelerations, which push some elements, especially the heavy ones, toward the surface. \cite{Valle2014, Valle2015} found that neglecting atomic diffusion can lead to uncertainties in the mass, radius, and age of 4.5\%, 2.2\%, and 20\%, respectively. Using a sample of \textit{Kepler} stars with masses lower than $M=1.2~\Msun$, \cite{Nsamba2018} found systematic differences in age of up to 16\%. \cite{Cunha2021} performed a model-based control study and found that the impact on the inference of stellar age is about 10\% for a 1.0~M$_\odot$ star near the end of the main sequence.

The study of atomic diffusion is key to reducing uncertainties in stellar characterization. However, radiative accelerations are often ignored. This process is, in fact, extremely computationally demanding because it requires a deep understanding of opacities (usually requiring the inclusion of monochromatic opacity). While gravitational settling reproduces surface abundances for low-mass stars \citep{Chaboyer2001,Salaris2001}, radiative accelerations are required for stars with thin convective envelopes (e.g., for solar-metallicity stars with effective temperatures higher than $\sim6000$~K; \citealt{michaud15}).  Recently, \cite{Rehm2024} showed that chemical accumulation zones caused by radiative accelerations in B-type stars can excite previously missing modes   .

\cite{Seaton1997} proposed an interpolation method that was integrated into the OP monochromatic opacity package developed by \cite{Seaton2005} to determine the necessary parameters for the calculation of radiative accelerations. This method has been implemented in several stellar evolution codes, for example TGEC \citep{Theado2009,huibonhoa24} and Modules for Experiments in Stellar Astrophysics \citep[MESA;][]{Paxton2013}, following the method outlined in \cite{Hu2011}. Recent work by \cite{Mombarg2022} optimized the \cite{Seaton1997}  method by computing only the necessary parameters when the stellar model's internal structure changes significantly. The \cite{Mombarg2022} method has been implemented in recent versions of MESA \citep{Jermyn2023}. 
\cite{LeBlanc2004} and \cite{alecian20} developed a parametric approximation to compute radiative accelerations, avoiding the on-the-fly consideration of monochromatic opacities: the single-valued parameters (SVP) method. This method separates the dependence on atomic data and chemical composition using an analytical expression for radiative accelerations. This method is slightly less accurate than the previous ones (with at most 30\% uncertainties) but enables much faster computations ($10^3$ times faster than the \citealt{Seaton1997} method). Comparison between the \cite{Seaton2005} and SVP methods was performed by \cite{campilho22} and \cite{huibonhoa24}, who find that the two methods predict very similar surface abundances.

In \citet[ hereafter Paper I]{moedas2022} we presented another approach to calibrating radiative accelerations in a turbulent diffusion coefficient: a formulation that parameterizes chemical transport mechanisms in competition with atomic diffusion. Turbulent mixing is needed for F-type and more massive stars to avoid unrealistic surface variations caused by atomic diffusion \citep{Verma2019}.
In Paper I  we found that it is necessary to increase the efficiency of turbulent mixing with the stellar mass to reproduce the effects of radiative acceleration on iron. It improves stellar characterization and allows atomic diffusion to be included in stellar models without producing unrealistic abundances variations. We showed in \citet[ hereafter Paper II]{Moedas2024} that neglecting atomic diffusion can lead to age uncertainties of more than 20\% for F-type stars. However, this prescription cannot reproduce the full chemical evolution in stellar models. It is in fact valid only for iron, the element used to constrain the approximation, but not for other elements (e.g., calcium accumulation at the stellar surface; see Paper I). 

In this work we studied the SVP method to include radiative accelerations in the stellar models, which allowed us to improve chemical predictions without any great loss of efficiency. We implemented the SVP method in the r12778 version of MESA to study how it affects the stellar models and how it performs in characterizing a sample of \textit{Kepler} observed stars.


This article is structured as follows. In Sect. \ref{sec:Stellar_Models} we present the input physics of the grids of stellar models. In Sect. \ref{sec:SVP} we present the SVP method and how it affects the stellar models. The stellar sample, the optimization process, the grids of stellar models we used, and the results for the fundamental properties are presented in Sect. \ref{sec:imp_SVP}. In Sect. \ref{sec:discussion} we discuss the impact of the different methods on the surface abundances. We conclude in Sect.~\ref{sec:conclusion}.

\section{Stellar models}
\label{sec:Stellar_Models}

Stellar models were computed with the MESA r12778 evolutionary code \citep{Paxton2011, Paxton2013, Paxton2015, Paxton2018, Paxton2019}. 
This version includes a routine for calculating radiative accelerations that follow the work of \cite{Hu2011}. However, the present method is not efficient enough to compute the large number of stellar models necessary to characterize stars by grid-base modeling methods. For faster computation, we implemented the SVP method in this version of MESA. This method allows the use of simpler opacity tables for the computation of the Rosseland mean opacity, which significantly speeds up the computation. The use of these opacity tables is appropriate if the metal mixture is not strongly modified (e.g., the OPAL table; \citealt{Iglesias1996}). This approximation of the computation of the Rosseland mean is justified by the fact we also included a turbulent diffusion coefficient that reduces the change of metal mixture for the stars analyzed in this work (see Sect.~\ref{sec:SVP}). The detailed steps of the implementation of the SVP method in MESA are presented in Appendix \ref{ap:how_SVP_implementation}. {We note that we did not test the new method implemented in MESA \citep{Mombarg2022,Jermyn2023} since it is not available in the version of the code we used, to be consistent with the one we used in Papers I and II. Further comparisons will be performed in the future.}

\subsection{Input physics}
\label{sec:Stellar_Physics}

In this work we adopted the following input physics.
We included the solar heavy elements mixture given by \citet{Asplund2009}, and the OPAL\footnote{\url{https://opalopacity.llnl.gov/}} opacity tables \citep{Iglesias1996} for the higher-temperature regime, and the tables provided by \cite{Ferguson2005} for lower temperatures.  All tables are computed for a given mixture of metals. We adopted the OPAL2005 equation of state \citep{Rogers2002}. We used the \cite{Krishna1966} atmosphere for the boundary condition at the stellar surface. We followed the \cite{Cox1968} for convection, imposing the mixing length parameter ($\alpha_\mathrm{MLT}$) in agreement with the solar calibration we performed for the grids. In the presence of a convective core, we implemented core overshoot following an exponential decay with a diffusion coefficient, as presented in \citet{Herwig2000}:  
\begin{equation}
    D_{\rm ov}=D_0\exp\left(-\frac{2z}{fH_p}\right),
\end{equation}
where $D_0$ is the diffusion coefficient at the border of the convective unstable region, $z$ is the distance from the boundary of the convective region, $H_p$ is the pressure scale height, and $f$ is the overshoot parameter. {In MESA, it is necessary to define an additional parameter $f_0$, the starting point of the overshoot mixing. We set $f_0=0.0001$ and $f=0.0101$ to have an overshoot of 0.01.} 

\subsection{Turbulent mixing}
Here we present a brief description of turbulent mixing prescriptions implemented in the models. More detail are provided in Paper I and references therein. We included turbulent mixing in two grids of this work, following two prescriptions. One is the prescription proposed by \cite{Proffitt1991},
\begin{equation}
    D_\mathrm{T}=C\left(\frac{\rho_\mathrm{BC}}{\rho}\right)^n,
    \label{eq:Dturb}
\end{equation}
where $C$ and $n$ are constants, $\rho$ is the local density, and the  $\rho_\mathrm{BC}$ is the density at the bottom of the convective zone.
The other prescription is taken from \cite{Richer2000},
\begin{equation}
    D_\mathrm{T}=\omega D(\rm{He})_0\left(\frac{\rho_0}{\rho}\right)^n,
    \label{eq:Dturb2}
\end{equation}
where $\omega$ and $n$ are constants, $D(\rm{He})$ is the local diffusion coefficient of helium, and the index 0 indicates that the value is taken at a reference depth. The $D(\rm{He})$ is computed following the analytical expression given by \cite{Richer2000}:
\begin{equation}
    D(\mathrm{He})=\frac{3.3\times10^{-15}T^{2.5}}{4\rho\ln{(1+1.{1}25\times10^{-16}T^3/\rho)}},
    \label{eq:DHe}
\end{equation}
where $T$ is the local temperature. In this work we set $\omega$ and $n$ to $10^4$ and $4$, respectively \citep{Michaud2011a,Michaud2011b}.

For one grid, we considered the turbulent mixing parameterization presented in Paper I, where we added the effects of radiative acceleration on iron.
We used a reference envelope mass ($\Delta M_0$, as the reference depth) to indicate the depth that turbulent mixing reaches inside the star. We showed that this parameter varies with the mass of the star as  
\begin{equation}\label{eq:param}
    \Delta M_0 \left(\frac{M^\ast}{M_\odot}\right)= 3.1\times 10^{-4} \times \left(\frac{M^\ast}{M_\odot}\right) + 2.7\times 10^{-4}.
\end{equation}
Higher values of $\Delta M_0$ result in more efficient mixing due to turbulent mixing, which in turn accounts for the radiative acceleration process. Models with our calibrated turbulent mixing we refer to them as~$\Dtfe$. 

{When we considered the individual effect of radiative accelerations, we also adopted the turbulent mixing coefficient calibrated by \citet[ hereafter VSA19]{Verma2019}, where $\Delta M=10^{-4}~\Msun$. This allowed us to avoid unrealistic chemical surface variations without introducing the extra effect of radiative accelerations in the ~$\Dtfe$. }

{We used both turbulent mixing prescriptions for the models that included the SVP method for radiative accelerations depending on the mass of the type of stars. For GK-type stars, we used the \cite{Proffitt1991} prescription (Eq. \ref{eq:Dturb}), which allows us to reproduce the lithium abundance of the Sun ($C=1615.4763$ and $n=1.3$ from the solar calibration). For F-type stars, we considered the prescription proposed by \citet[ see Eq. \ref{eq:Dturb2}]{Richer2000} using the calibration done by VSA19, which allowed us to avoid the unrealistic surface abundance variations caused by atomic diffusion). 
To determine which prescription to use, we considered the size of the convective shell mass ($M_\mathrm{CZ}$) at the zero-age main sequence. If $M_\mathrm{CZ}\geq10^{-4}\Msun$, we used \cite{Proffitt1991}; otherwise, we considered the other. We note that the efficiency of turbulent mixing prescription derived by \cite{Proffitt1991} depends on the size of the convective envelope, allowing a smooth transition between the two prescriptions with this criterion.}

\section{Modeling of radiative accelerations}
\label{sec:SVP}

{In this section we investigate how the SVP method affects the stellar models, comparing it to the method of \cite{Hu2011} and the method we presented in Paper I. We computed individual stellar models for each method.}

\subsection{The single-valued parameters method}

The SVP method \citep{LeBlanc2004, alecian20} separates the dependence on atomic data and element abundances. It then parameterizes the atomic data into six parameters whose values are provided with precalculated tables. This parameterization is made possible by the fact that some quantities are almost independent of frequencies for the given stellar interior conditions. The routines to compute radiative accelerations with SVP {(version released in 2020)} are publicly available\footnote{\url{https://gradsvp.obspm.fr/index.html}}. These routines require a few input parameters, including temperature, density, pressure, and mass ($T$, $\rho$, $P$, $L$, $r$, and $m$). They also require the mass fraction of the followed elements, $X_\mathrm{i}$. The method is currently only applicable to main-sequence stars, and models with masses between $1.0$ and $10~\Msun$ (as the tables are only prepared for this stage). The implementation in MESA is complicated by the fact that the code does not provide an external hook\footnote{external routines that can be modified without changing the source code.} for the radiative accelerations. In this case, it is necessary to modify the MESA source code directly. To minimize code changes, we added a new hook to MESA to be used as an alternative to the default radiative accelerations. The implementation steps are described in Appendix~\ref{ap:how_SVP_implementation} and a schematic is shown in Fig.~\ref{fig:SVP_imp_sch}.

\begin{figure}[]
\centering
  \centering
  \includegraphics[width=0.9\columnwidth]{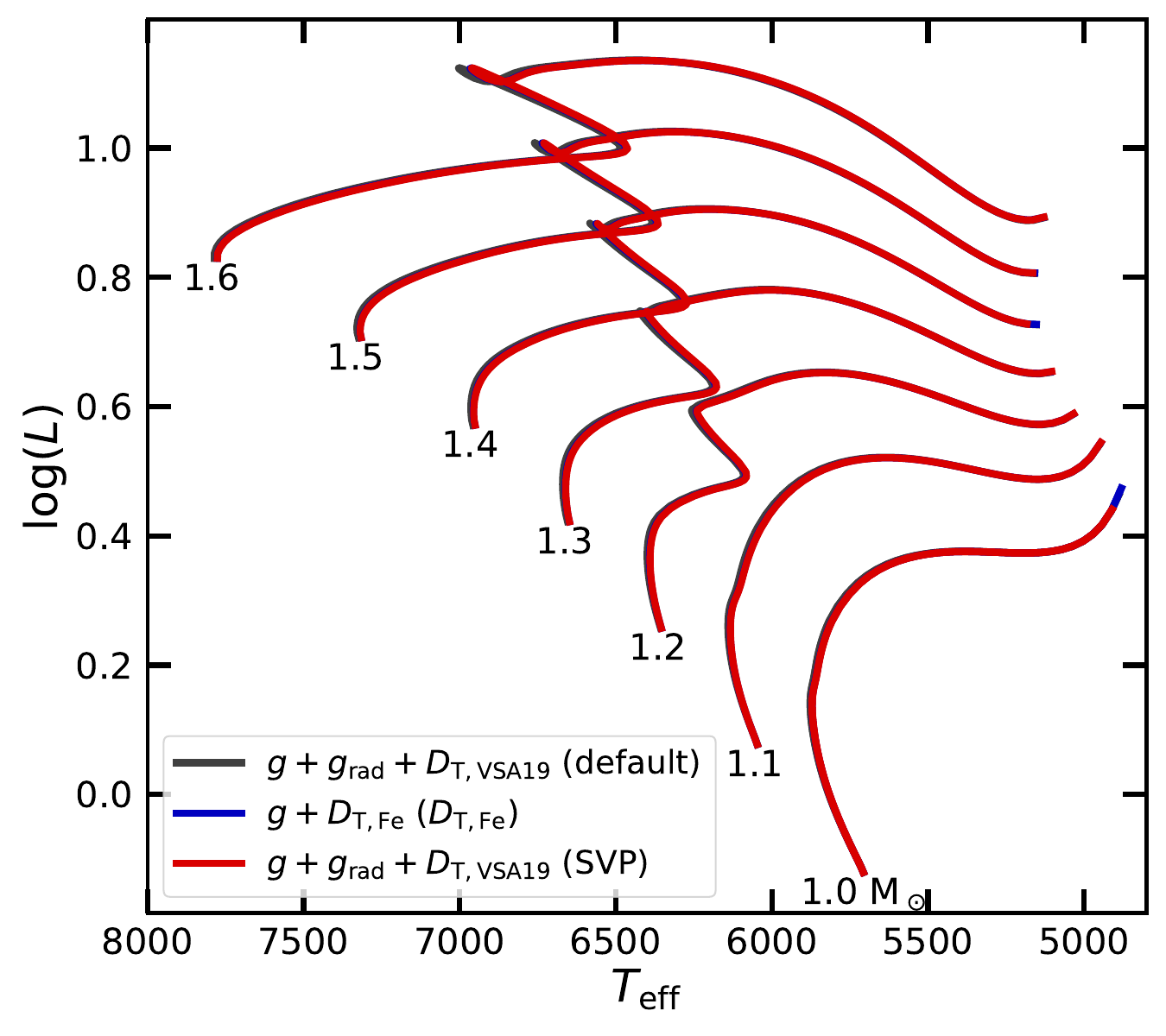}
  \includegraphics[width=0.9\columnwidth]{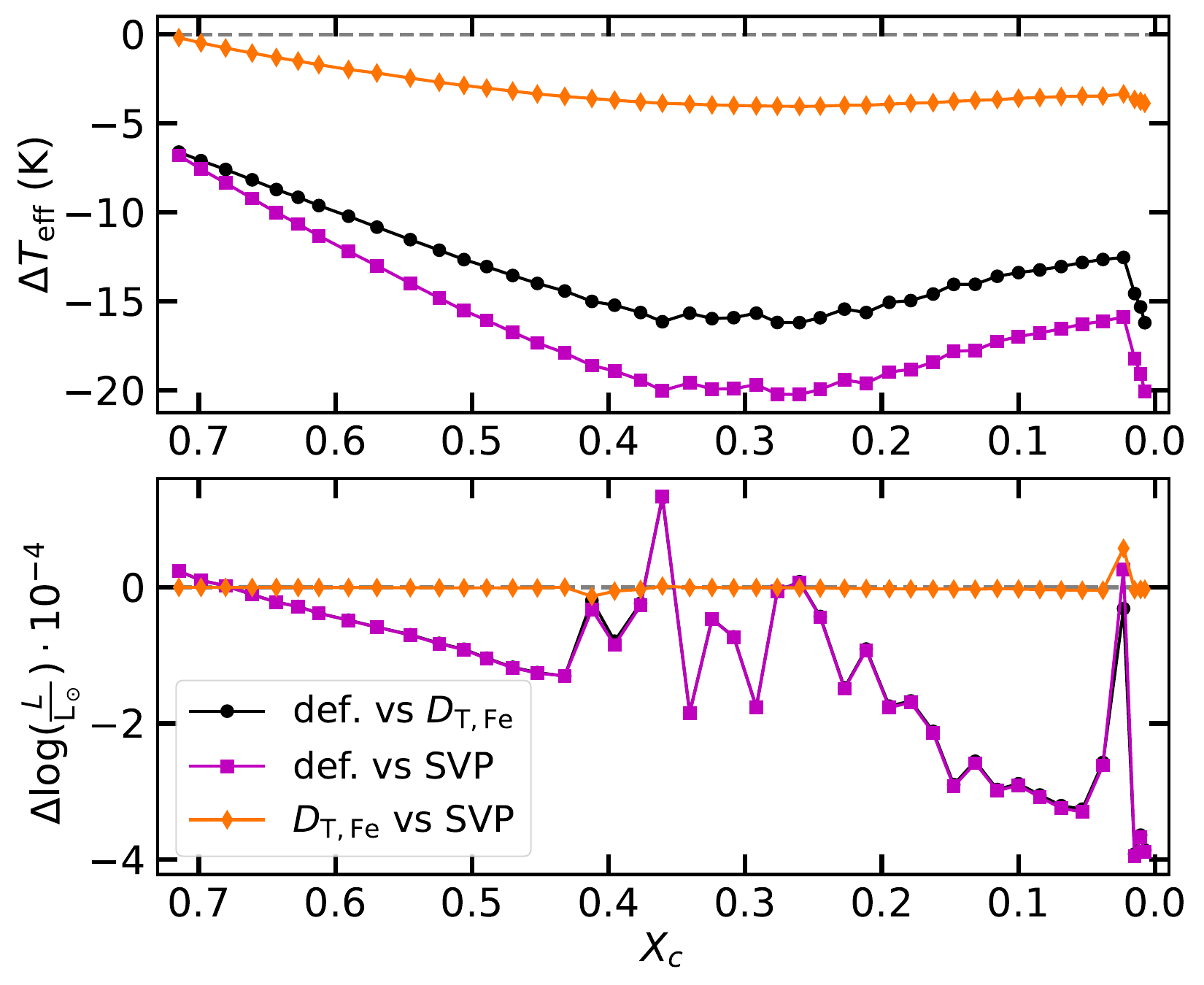}
\caption{Upper panel: HR diagram of different evolutionary tracks for the different radiative acceleration methods -- the default method (gray line), models with $\Dtfe$ (blue line), and the SVP method (red line). Lower panels: Difference for the 1.4 $\Msun$ models for the different methods, for effective temperature (middle panel) and luminosity (bottom panel) as a function of the central hydrogen mass fraction. The black circles are the differences between the default method and the $\Dtfe$, the purple squares the differences between the default method and the SVP method, and the orange diamonds the differences between $\Dtfe$ and the SVP method.}
    \label{fig:HR_diff}
\end{figure}

\begin{figure*}[]
    \centering
    \includegraphics[width=0.9\textwidth]{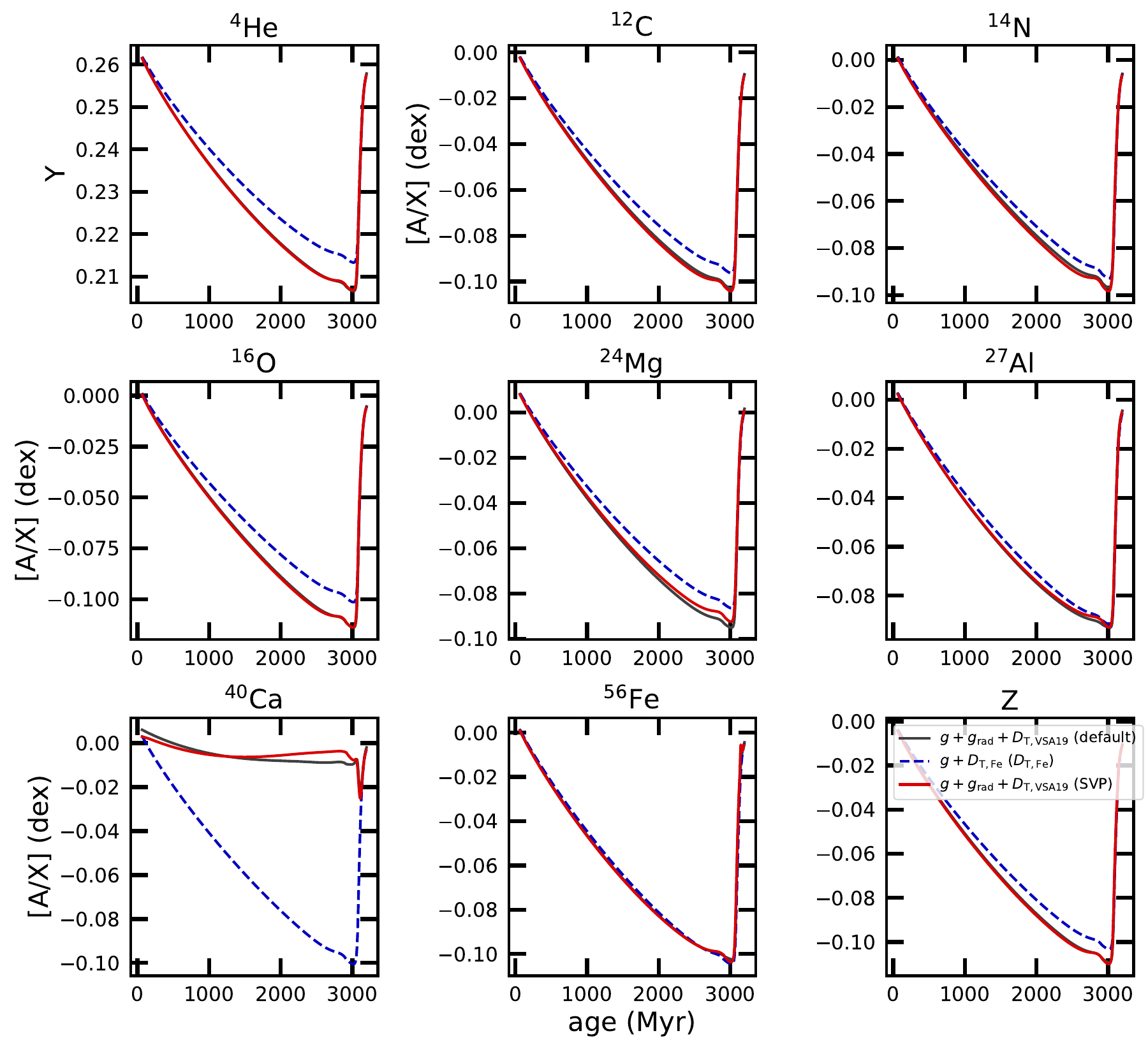}
    \caption{ Evolution of the surface abundance of some chemical elements for a 1.4 $\Msun$ model. The solid gray lines represent a default model, the dashed blue lines the model with $\Dtfe$ (Paper I), and the dot-dashed red  lines the the model that considers the SVP method.}
    \label{fig:SVP_elm_ev}
\end{figure*}

\subsection{Impact of radiative accelerations on stellar models}

We computed seven evolutionary tracks for each method. The first includes radiative accelerations computed with the \cite{Hu2011} method, and also VSA19 turbulent mixing parametrization to avoid unrealistic surface abundance variations (we refer to it as the {default} method). The second uses the models that include the turbulent mixing calibration presented in Paper I (we refer to it as the $\Dtfe$ method). The last one uses the SVP method and includes the {VSA19 turbulent mixing calibration ($\DtV$)} similar to the default model (we refer to it as the SVP method).

Figure \ref{fig:HR_diff} shows the models in the Hertzsprung-Russell (HR) diagram (top panel) with the different ways of accounting for radiative accelerations.
The two bottom panels show the difference between $\Teff$ and $\log(L)$ for the 1.4~$\Msun$ model with the default method as reference (black and purple circles) and the calibrated turbulent mixing as reference (orange circles). The three types of models almost overlap in the HR diagram, indicating that the processes do not have a significant effect on this aspect of stellar evolution. This is supported by the fact that the differences shown in the HR diagram are smaller than the uncertainties in $\Teff$ and $L$ expected from observations.

Despite showing negligible differences in the HR diagram, there are two main differences between the methods. The first one is the computational cost, which can be very different from one method to another. Compared to the standard method, the SVP method is almost 20 times faster in computing the models (in this case from the zero-age main sequence to the end of the subgiant stage). Compared to the $\Dtfe$ method, the SVP method is 1.5 times slower, due to the treatment of radiative acceleration for each element. Nevertheless, the SVP method is more efficient than the standard method, which allows the effects of radiative acceleration to be taken into account in individual elements.

\begin{figure}[]
    \centering
    \includegraphics[width=0.95\columnwidth]{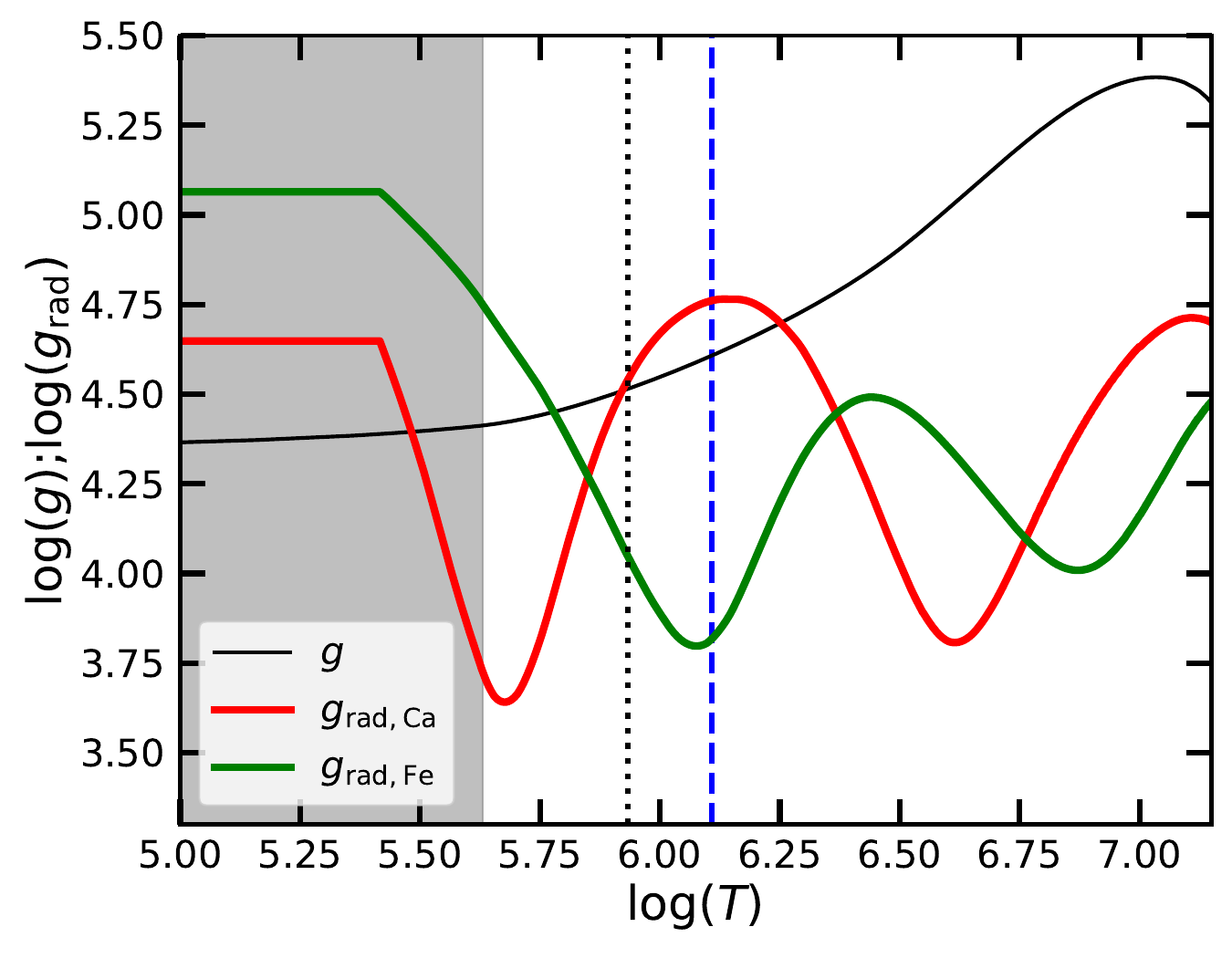}
    \caption{Profile for gravity (solid black line) and radiative accelerations for Ca (solid red line) and Fe (solid green line) for a 1.4~$\Msun$ model. The gray region is the convective envelope. The dotted black line indicates the reference envelope mass used for the turbulent mixing calibration from VSA19, and the dashed blue line is the reference envelope mass ($\Delta M_0=2.4\cdot10^{-3}~\Msun$) for the maximum accumulation of Ca at the surface of the model.}
    \label{fig:rad_profile2}
\end{figure}

The second major difference between the methods is their effect on chemical evolution. Figure \ref{fig:SVP_elm_ev} shows the surface abundance evolution of several elements using the different treatments of the transport. For iron, all methods reproduce similar evolution. For the other elements, the SVP method predicts consistent evolution with the default method. In the case of calcium, the $\Dtfe$ method cannot reproduce the effects of radiative acceleration, while the SVP method predicts a much closer evolution. 
{The behavior of calcium at the surface is due to the radiative accelerations. Figure~\ref{fig:rad_profile2} shows the radiative profiles of iron and calcium for the $1.4~\Msun$ model. In the case of the default and SVP method, the radiative acceleration is directly taken into account in the stellar model while for the $\Dtfe$ model the effect is included in a turbulent diffusion coefficient. When $\DtV$ (vertical dotted black line) is included, the mixing reaches a region where the radiative acceleration is higher than the gravitational settling, leading to a constant or an accumulation of the element. In the case of $\Dtfe$ there is no individual effect of radiative acceleration that leads to a depletion of calcium, as we see in Fig.~\ref{fig:SVP_elm_ev}.}  
As expected, this indicates that SVP allows for a better reproduction of the effects of radiative accelerations for each element in the stellar models compared to the calibration presented in Paper I, with a more efficient computation than the default method available in MESA.

\section{Impact of radiative acceleration on stellar fundamental properties}
\label{sec:imp_SVP}

{Next we investigated how the SVP method affects the inference of the fundamental properties of observed stars. We used a sample of \textit{Kepler} stars (Sect. \ref{sec:sample}) and inferred their properties  using three different grids (Sect. \ref{sec:grids})}  

\subsection{Stellar sample}
\label{sec:sample}

\begin{figure}
    \centering
    \includegraphics[width=0.93\columnwidth]{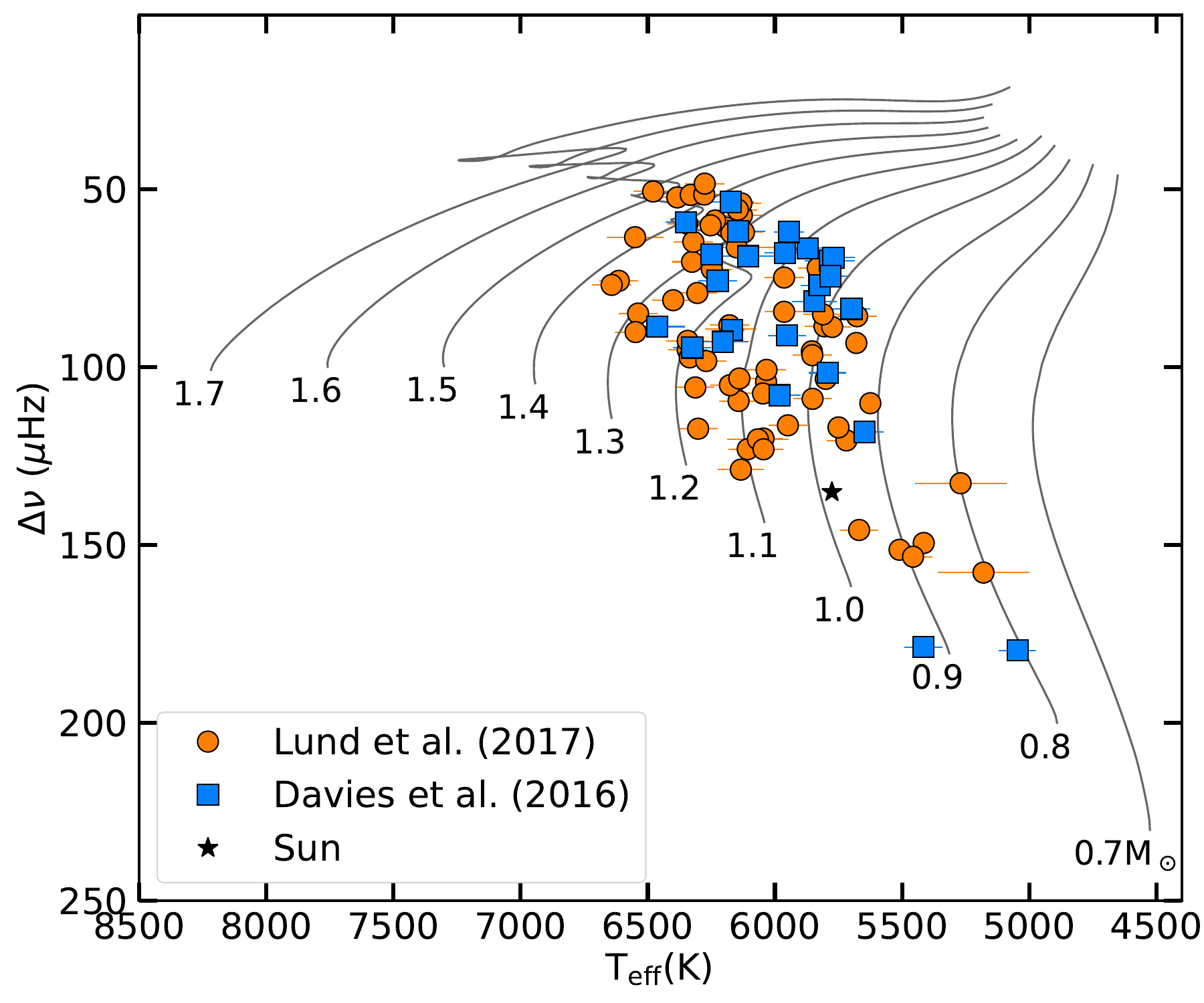}
    \caption{Asteroseismic diagram showing some computed evolutionary tracks with $\MH_i=0.0$ and $Y_i=0.26$ (which are not the solar calibrated values) in solid black lines. The symbols show the distribution of the sample considered in this work; orange circles are taken from \cite{Lund2017}, blue squares from \cite{Davies2016}, and the black star is the Sun.}
    \label{fig:sample}
\end{figure}

The sample of stars is the same as in Paper II, and it is selected from two different sources. The first source is the \textit{Kepler} LEGACY sample from \citet[][ hereafter L17]{Lund2017}. The second source is the work of \citet[][ hereafter D16]{Davies2016}, a study of 35 \textit{Kepler} stars (32 of them different from L17) that are planet hosts. For both samples, we only selected stars with $\FeH>-0.4$ to stay within the parameter space of the grids. The smallest value of $\FeH$ is -0.37~dex. We note that the value of $\FeH$ is usually smaller than $\MH$ in the stellar models, reaching a difference of up to 0.04~dex, as reported in the Paper~I, allowing the stars selected to be well within the grid parameter space. We excluded two stars from D16 that show mixed modes (KIC7199397 and KIC8684730). Finally, we included the degraded Sun as presented in \cite{Lund2017} as a control star. Thus, we obtained a sample of 91 stars (62 from L17, 28 from D16, and the Sun). The distribution of the full sample is presented in the asteroseismic diagram of Fig.~\ref{fig:sample}.
 
For all stars, we used the effective temperature ($\Teff$), iron content ($\FeH$), luminosity ($L$), and individual seismic frequencies ($\nu_{i}$) as constraints. We computed the $L$ from the magnitudes and parallaxes of \textit{Gaia} DR3 \citep{Gaia_Collaboration2021}. The other constraints were taken from the respective papers, except for 13 stars of L17, for which we used the $\Teff$ and $\FeH$ values updated in \cite{morel2021}.

\subsection{Optimization process}
\label{sec:opt}

The optimization process we adopted is the same as in Paper II, and we give here a brief overview. The fundamental properties of the sample are inferred with the asteroseismic inference on a massive scale (AIMS) tool (\cite{Rendle2019} code. AIMS is an optimization tool that uses Bayesian statistics and Markov chain Monte Carlo to explore the parameter space of the grids and find the model that best fits the observational constraints.
In the present work we used the two-term surface corrections proposed by \cite{Ball2014} to compensate for the difference between theoretical and observed frequencies due to incomplete modeling of the surface layers of stars.
AIMS distinguishes the contribution of the global constraints, $X_i$ (in our case $\Teff$, $\FeH,$ and $L$),
\begin{equation}
    \label{eq:Chi_class} 
    \chi^2_\mathrm{global}=\sum^3_i\left(\frac{X_i^\mathrm{(obs)}-X_i^\mathrm{(mod)}}{\sigma(X_i)}\right)
\end{equation}
and the constraints from individual frequencies, $\nu_{i}$,
\begin{equation}
    \label{eq:Chi_seis} 
    \chi^2_\mathrm{freq}=\sum^N_i\left(\frac{\nu_{i}^\mathrm{(obs)}-\nu_{i}^\mathrm{(mod)}}{\sigma(\nu_{i})}\right),
\end{equation}
where "(obs)" corresponds to the observed values and (mod) corresponds to the model values.
The weight that AIMS gives to the seismic contribution can be absolute (3:N), where each individual frequency has the same weight as each global constraint,
\begin{equation}
    \label{eq:Chi_tot_abs} 
    \chi^2_\mathrm{total}=\chi^2_\mathrm{freq}+\chi^2_\mathrm{global},
\end{equation}
or relative (3:3), where all the frequencies have the same weight as all the global constraints,
\begin{equation}
    \label{eq:Chi_tot_rel} 
    \chi^2_\mathrm{total}=\left(\frac{N_\mathrm{global}}{N_\mathrm{freq}}\right)\chi^2_\mathrm{freq}+\chi^2_\mathrm{global},
\end{equation}
where $N_\mathrm{global}$ and $N_\mathrm{freq}$ are the numbers of global and frequency constraints, respectively.

The absolute and relative weight impact the stellar characterization especially the inferred uncertainties. Nonetheless, in this work we only compare the results obtained with absolute weight in the frequencies as it does not affect the conclusion of this work.

\begin{table*}[]
\caption{Parameter space and additional input physics of the three grids. }
\label{tab:stellar_grids}
\centering
\resizebox{0.85\textwidth}{!}{%
\begin{tabular}{cccccccccc}
\hline
\multirow{2}{*}{Grid} & \multicolumn{2}{c}{Mass (M$_\odot$)} & \multicolumn{2}{c}{$\rm [M/H]_i$} & \multicolumn{2}{c}{$Y_i$} & \multicolumn{2}{c}{Atomic Diffusion} & Turbulent \\ \cline{2-9}
 & Range & Step & Range & Step & Range & Step & $g$ & $g_\mathrm{rad}$ & Mixing \begin{tabular}[c]{@{}c@{}}  \end{tabular} \\ \hline
A & \multirow{5}{*}{[0.7;1.75]} & \multirow{5}{*}{0.05} & \multirow{5}{*}{[-0.4;0.5]} & \multirow{5}{*}{0.05} & \multirow{5}{*}{[0.24;0.34]} & \multirow{5}{*}{0.01} & $\Delta \FeH_\mathrm{Max}<0.2$ & \multirow{3}{*}{No} & No \begin{tabular}[c]{@{}c@{}}\phantom{a}\\\phantom{a}\end{tabular}\\ \cline{1-1} \cline{8-8} \cline{10-10} 
B &  &  &  &  &  &  & \multirow{3}{*}{All Models} &  & $D_\mathrm{T,Fe}$ \begin{tabular}[c]{@{}c@{}}\phantom{a}\\\phantom{a}\end{tabular} \\ \cline{1-1} \cline{9-10} 
C &  &  &  &  &  &  &  & $M>1.0 \Msun$ &\begin{tabular}[c]{@{}c@{}}$D_{T,\mathrm{PM91}}$ if $M_\mathrm{CZ}\geq 10^{-5}~\Msun$\\ $D_{T,\mathrm{VSA19}}$ if $M_\mathrm{CZ}< 10^{-5}~\Msun$\end{tabular} \\ \hline
\end{tabular}%
}\tablefoot{$D_{T,\mathrm{Fe}}$ is the calibration of turbulent mixing presented in Paper I, $D_{T,\mathrm{VSA19}}$ is the calibration done by VSA19 (both based on the \citealt{Richer2000} prescription), and $D_{T,\mathrm{PM91}}$ is the prescription of \cite{Proffitt1991}.}
\end{table*}

\subsection{Parameter space of the grids}
\label{sec:grids}

To understand the effect of radiative accelerations on stellar fundamental properties, we built three grids of stellar models. The three grids cover the same parameter space in mass $M$, initial metallicity $\MH_\mathrm{i}${\footnote{$\MH=\log({Z}/{X})-\log({{Z}/{X}})_\odot$}} and the initial fractional abundance of helium $Y_i$ (see Table \ref{tab:stellar_grids}). The main difference is in the chemical transport mechanisms included in the models. Two of the three grids (grids A and B) are the same as those presented in Paper II.

Grid A only includes atomic diffusion without radiative accelerations. In this case, to avoid the effects of unrealistic surface abundances this process is turned off for F-type and more massive stars. We used the $\Delta\FeH_\mathrm{Max}$\footnote{$\Delta\FeH=\mid\FeH-\FeH_i\mid$}$>0.2$~dex to define the evolutionary tracks that present an extreme surface abundance variation. This criteria takes into consideration the effects of different mass and chemical compositions on the surface abundance variation  (see Paper I for more details).


Grid B includes atomic diffusion (without radiative accelerations) and the turbulent mixing parametrization $(\Dtfe)$ presented in Paper I, where the efficiency increases with stellar mass according to Eq.~\ref{eq:param}. This prescription avoids unrealistic surface abundance variations, allowing atomic diffusion to be included in all grids of stellar models. Also, the calibration provided in Paper I adds the global effect of radiative accelerations on iron. Grids A and B have the same $\alpha_\mathrm{MLT}=1.711$ from their solar calibration.

Grid C includes the SVP method for calculating radiative accelerations. However, the SVP tables are valid for stars with masses between $1.0$ and $10.0~\Msun$. Hence, in this grid only stellar models with masses $\geq1.0~\Msun$ include radiative accelerations. We also included turbulent mixing in the stellar models, considering two prescriptions in different regimes.

\begin{figure*}[]
    \centering
    \includegraphics[width=1\textwidth]{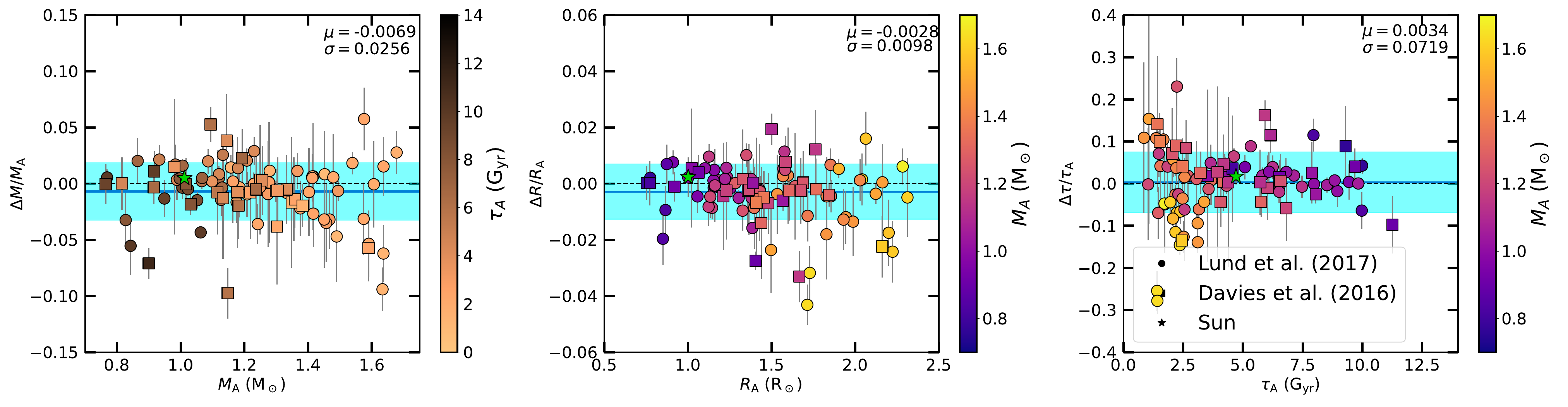}
    \caption{Relative difference for mass (left panel), radius (middle panel), and age (right panel) between grids A and C. The solid blue line indicates the bias, and the blue region is the 1$\sigma$ of the standard deviation. Each point is color-coded with the corresponding reference age (left panel) and mass (middle and right panels).}
    \label{fig:stellar_prop_SVP_nodiff}
\end{figure*}

\subsection{Results}
\label{sec:results}
To understand the impact of the different input physics of the grids we investigated the absolute differences 
\begin{equation}
    \Delta X = X_C - X_r,
\end{equation}
and the relative difference with
\begin{equation}
    \frac{\Delta X}{X_r},
\end{equation}
where X is the inferred parameter value, $X_C$ is the value obtained from grid C, and $X_r$ is the reference value obtained whether grid A or B is used.

\begin{figure*}[]
    \centering
    \includegraphics[width=1\textwidth]{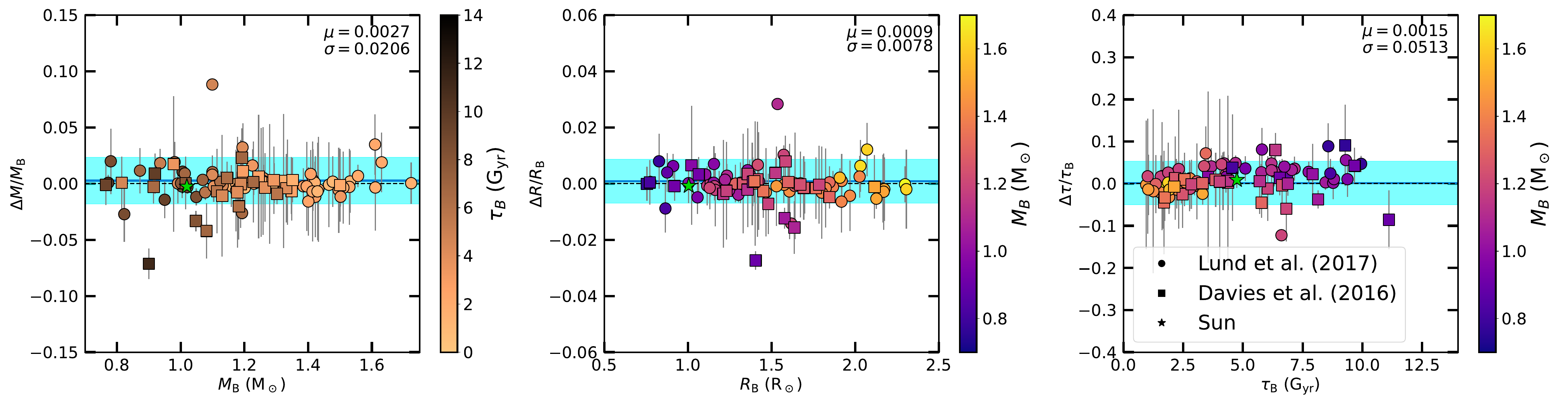}
    \caption{Same as Fig.\ 5 but for the differences between grids B and C. }
    \label{fig:stellar_prop_SVP_Dturb}
\end{figure*}

\subsubsection{Grid C versus grid A}
\label{sec:gridC_A}

We first compared the results obtained with grids A and C. Figure \ref{fig:stellar_prop_SVP_nodiff} shows the relative difference in the fundamental properties between the two grids. The results show a small bias, smaller than 1\%, and a dispersion of about 2.6\%, 0.9\%, and 7.2\% for mass, radius, and age.
{Compared to the individual error of the relative differences determined through the propagation of uncertainties (assuming the inferred results are independent), the biases ($\mu$) are two times smaller than the relative errors (values of the error in $\frac{\Delta X}{X_r}$ are larger than 0.015, 0.009, and 0.06 for mass, radius, and age, respectively.)}
This suggests that these properties are almost not affected by the inclusion of atomic diffusion with radiative accelerations. However, when we look at individual stars, this difference can be larger than 25\% for age. These results are in agreement with Paper II when comparing results between grids A and B. It is also possible to observe an increase in the dispersion (especially in the age) toward higher masses.
{ If we focus on stars with masses greater than $1.2\Msun$, the bias toward lower ages increases to 3\%.} This dispersion and bias in age are mainly caused by the fact that atomic diffusion is turned off in the models of grid A in this mass regime.
{Models with atomic diffusion evolves faster because the settling of heavier elements reduces the quantity of hydrogen in the core.} However, grid C shows a slightly larger relative difference compared to Paper II, with about 10\%, 4\%, and 29\% for mass, radius, and age, due to the inclusion of the individual effects of radiative accelerations with the SVP method.

For the lower masses, we obtained a higher dispersion at this regime compared to Paper II. This is not due to the inclusion of radiative accelerations as their effects are negligible at these masses (also they are not included for models with masses lower than $1.0~\Msun$) but due to the inclusion of a different prescription of turbulent mixing. In fact, grid C in this regime includes the prescription from \cite{Proffitt1991} calibrated to reproduce the lithium abundance at the surface.  This can lead to relative uncertainties of up to 8\%, 3\%, and 10\% for the determination of mass, radius, and age, respectively.

\subsubsection{Grid C versus grid B}
\label{sec:gridC_B}

The comparison of the estimated mass, radius, and age using grids B and C for the full sample is shown in Figs.~\ref{fig:stellar_prop_SVP_Dturb}. Overall, the results show very small biases (less than 1\% for each parameter) and a scatter of 2\%, 0.7\%, and 5\% for mass, radius, and age, respectively. {Compared to the error of the relative differences, the biases ($\mu$) are more than two times smaller than the relative errors (values of the error in $\frac{\Delta X}{X_r}$ are larger than 0.015, 0.009, and 0.05 for mass, radius, and age, respectively.)} This indicates that there is no significant effect of modeling the effect of radiative accelerations using a calibrated turbulent mixing coefficient or the SVP method for the inference of the fundamental properties of stars. This is explained by the fact that only iron is used to constrain the chemical composition. We also find a large scatter for the lower-mass stars for the same reasons as the previous section (different turbulent mixing prescription).

\section{Surface abundances}
\label{sec:discussion}

The main effect of using either the SVP method or a turbulent mixing calibration to model radiative accelerations is visible with the evolution of elements other than iron. The chemical abundances inferred from the different grids are hence not the same. We focused only on the surface abundances determined with grids B and C since these two grids include a more accurate modeling of the transport of chemical elements.

\citet[ hereafter M21]{morel2021} provides the surface abundances of carbon, magnesium, aluminum, sodium, silicon, calcium, and lithium for 13 stars, 12 of which are part of the sample studied here. This allows us to compare these observed abundances with the inferred ones from the grids.

\begin{figure*}[]
    \centering
    \includegraphics[trim=3cm 0 0 1.5cm, clip, width=.9\textwidth]{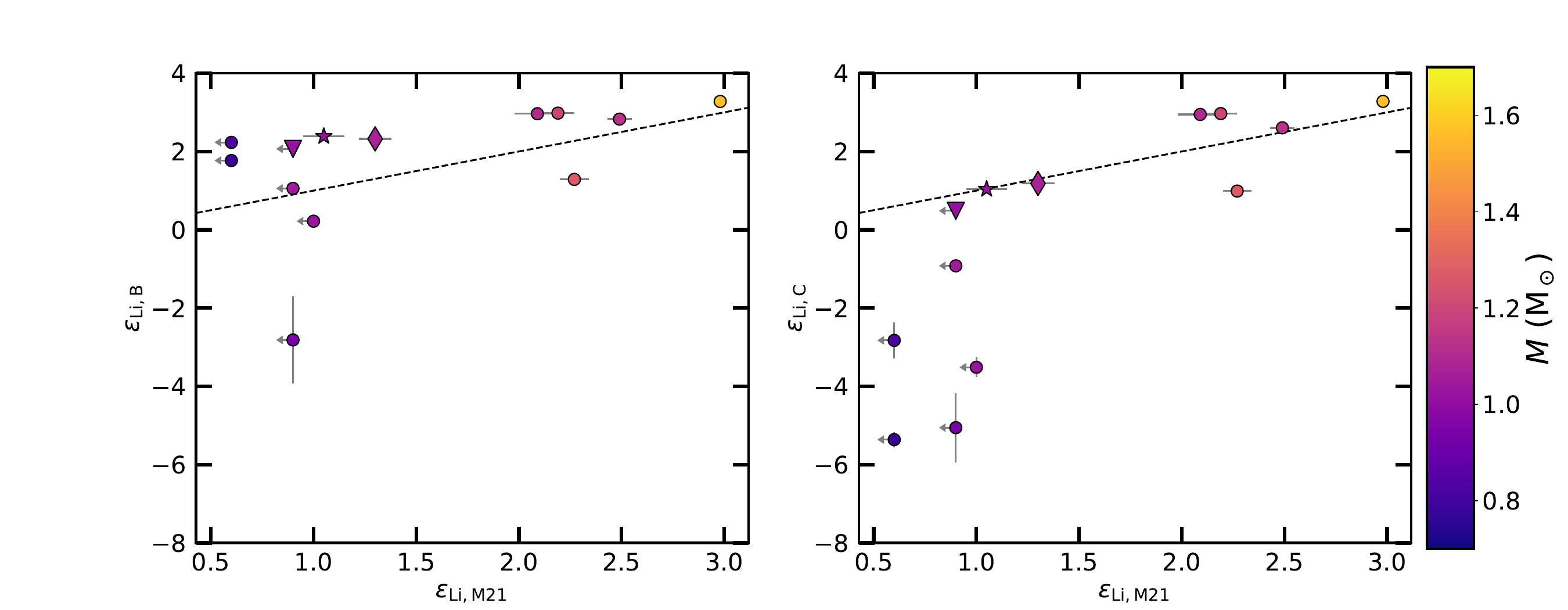}  
    \caption{Estimated lithium abundances versus observed abundances from M21 using grid B (left) and grid C (right). The diamonds and downward triangles refer to the 16 Cyg A and B stars (KIC~12069424 and KIC~12069449), respectively, and the star represents the Sun. The arrows pointing to the left indicate upper limits. The dashed line marks where the observed value is equal to the estimated value.}
    \label{fig:lithium}
\end{figure*}

\subsection{Lithium}
\label{sec:lithium}

Lithium is an important element for calibrating the extra mixing below the convective envelope because it is easily destroyed at low temperatures, providing strong constraints on the efficiency of the transport. For a standard solar model without extra mixing, the estimated abundance of lithium is much higher than the one determined from observations.
The lithium abundances estimated using the models including the turbulent mixing parametrization presented in Paper II are shown in the left panel of Fig. \ref{fig:lithium}. Most of the low-mass stars have higher lithium abundances than observed, indicating that the mixing in these models is not efficient enough to transport lithium into deeper layers where it is destroyed. The models with the calibrated turbulent mixing coefficient $D_\mathrm{T,Fe}$ cannot reproduce the observed abundances of lithium.

As expected, when turbulent mixing is calibrated to reproduce the lithium abundance of the Sun (grid C, right panel of Fig. \ref{fig:lithium}), the abundances of solar analogs are in better agreement with the determination of M21. This is confirmed with 16 Cyg A and B (KIC~12069424 and KIC~12069449), two solar-type stars that are usually used to check the quality of the solar-analog models. For 16 Cyg A, the inferred value is within 2$\sigma$ of the observed one, while for 16 Cyg B is in agreement with the observation within 1$\sigma$. It is important to keep in mind that the calibration of turbulent mixing for grid C was done with one star (the Sun), which is not sufficient since the mixing efficiency may be different for each star. This is the reason why for stars with masses lower than the Sun, the parametrization induces a depleting that is too large indicating it is not adapted. Moreover, additional events may affect the abundance of lithium such as the accretion of planetary matter during its evolution (e.g., \citealt{Deal2015} and references therein). 

For the higher-mass stars, there is no difference in the predictions because the radiative acceleration does not affect this element. Also, the difference in turbulent mixing efficiencies between the two grids is not large enough to produce a significant difference in the lithium surface abundances.

\subsection{Other elements}

\begin{figure*}[]
\centering
  \includegraphics[width=0.8\linewidth]{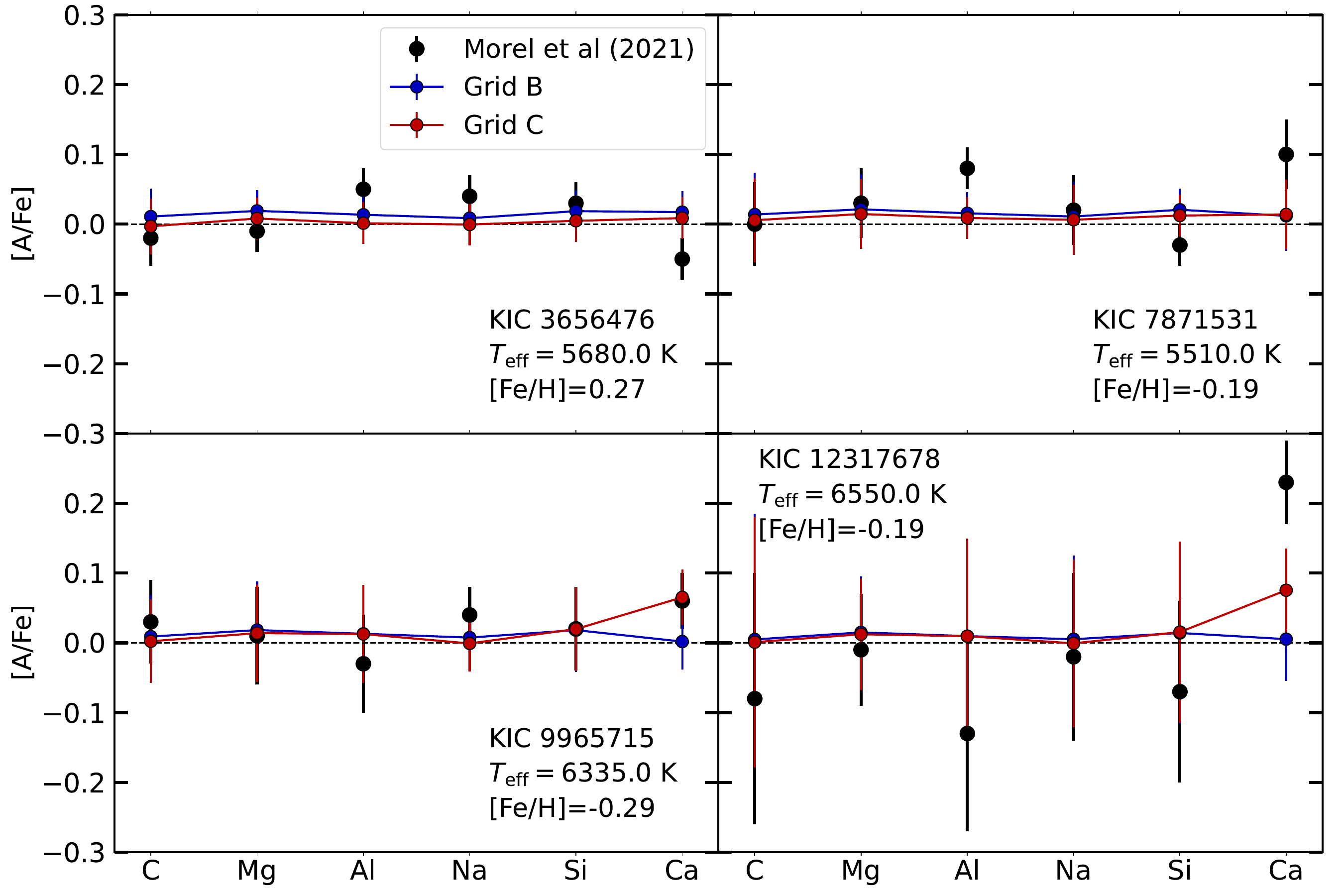}
\caption{Comparison of the estimated and observed abundances of four stars from the M21 sample.}
    \label{fig:obs_est_ele}
\end{figure*}

For the other elements, Fig. \ref{fig:obs_est_ele} shows the observed and predicted abundances (with grids B and C) for carbon, magnesium, aluminum, sodium, silicon, and calcium for four stars. These four stars are representative of the different results we obtained for the 13 stars. For the G-type stars (the upper panels), the predicted surface abundance of KIC~3656476 (left panel) is consistent with the observations for all elements. For KIC~7871531 (right panel) all predicted abundances are in agreement with the observation except for Al and Ca. This could indicate a slight $\alpha$ enrichment for this star, although there is no enrichment in Mg and Si (also $\alpha$ elements). If this is an $\alpha$-enriched star, our models are expected to not be able to reproduce these abundances as we do not change the element mixture (we only consider the solar one). For these two stars, there is no difference between the predictions obtained with grid B and C, as expected for G-type stars. 

For the two F-type stars (lower panels), the two grids yield similar abundances, except for calcium, which is higher when the SVP method is taken into account (grid C). Regarding the comparison with observations, for KIC~12317678 (right panel), the predicted surface abundances are in agreement with the observed ones, except for calcium. This star shows a high calcium abundance, and SVP models can predict abundances closer than turbulent mixing models. However, it is not sufficient to explain the observed surface abundance for this star. This indicates that either the efficiency of the extra mixing is not adapted (for calcium) in grid C or the star is enhanced in calcium compared to the Sun.

To better predict the surface [Ca/Fe] of KIC~12317678 the efficiency of turbulent mixing needs to be increased. It is important to note that for most of the elements, increasing the efficiency of the extra mixing decreases the impact of atomic diffusion. However, calcium is an exception in this case. Figure \ref{fig:rad_profile2} shows the radiative profiles of iron and calcium. In the case of calcium, enhancing the efficiency of turbulent mixing (by increasing the $\Delta M_0$), the mixing reaches a point where the radiative accelerations are higher than gravity. The maximum effects of radiative accelerations are observed around a reference mass envelope of $\Delta M_0=2.4\cdot10^{-3}~\Msun$ (blue dash line of the plot).

\subsection{Calcium abundance in F-type stars}

\begin{figure}[]
\centering
  \centering
  \includegraphics[width=0.95\columnwidth]{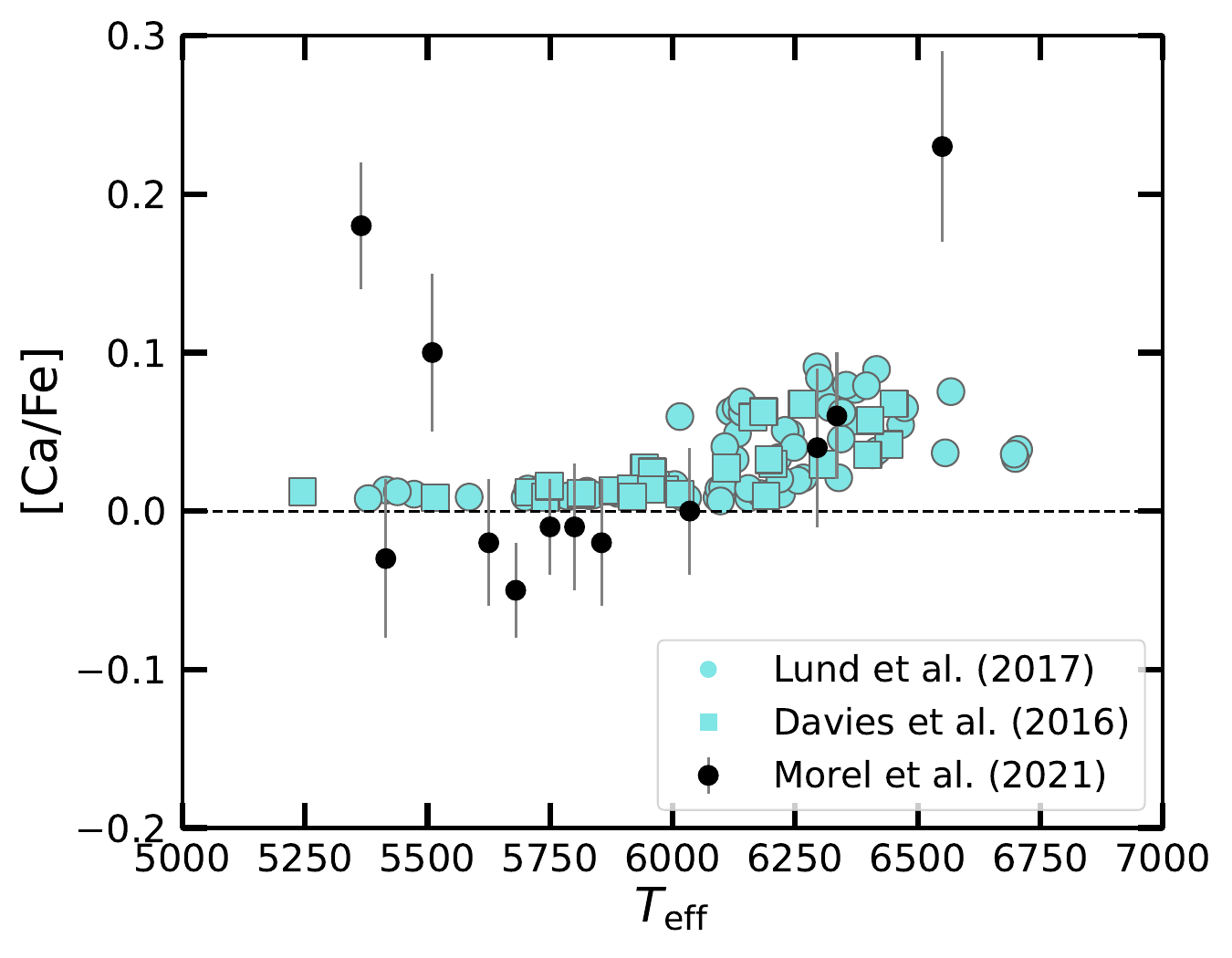}
\caption{Surface abundance of calcium. The blue symbols are the abundances predicted using grid C and the black symbols the observed abundances provided by \cite{morel2021}.}
    \label{fig:Ca_abun_est_vs_obs}
\end{figure}

Grid C includes a more realistic treatment of chemical transport mechanisms in the stellar models. If we look for the predicted surface [Ca/Fe] in our sample (see Fig. \ref{fig:Ca_abun_est_vs_obs}) there is an increase in its abundance for F-type stars as the $T_\mathrm{eff}$ increases (for stars with $T_\mathrm{eff}>6000K$). Compared to the abundances provided by M21, it seems that the prediction of the models is in agreement with observations, supporting the conclusion that radiative accelerations have a strong impact on calcium for F-type stars. 
There are also two low-temperature stars with higher calcium abundances, which are probably $\alpha$-enhanced stars, as they also show higher abundances of Mg, Al, and Si. However, since this is a small sample, there are no statistical arguments to support it. 

\begin{figure}[]
\centering
\includegraphics[width=0.95\columnwidth]{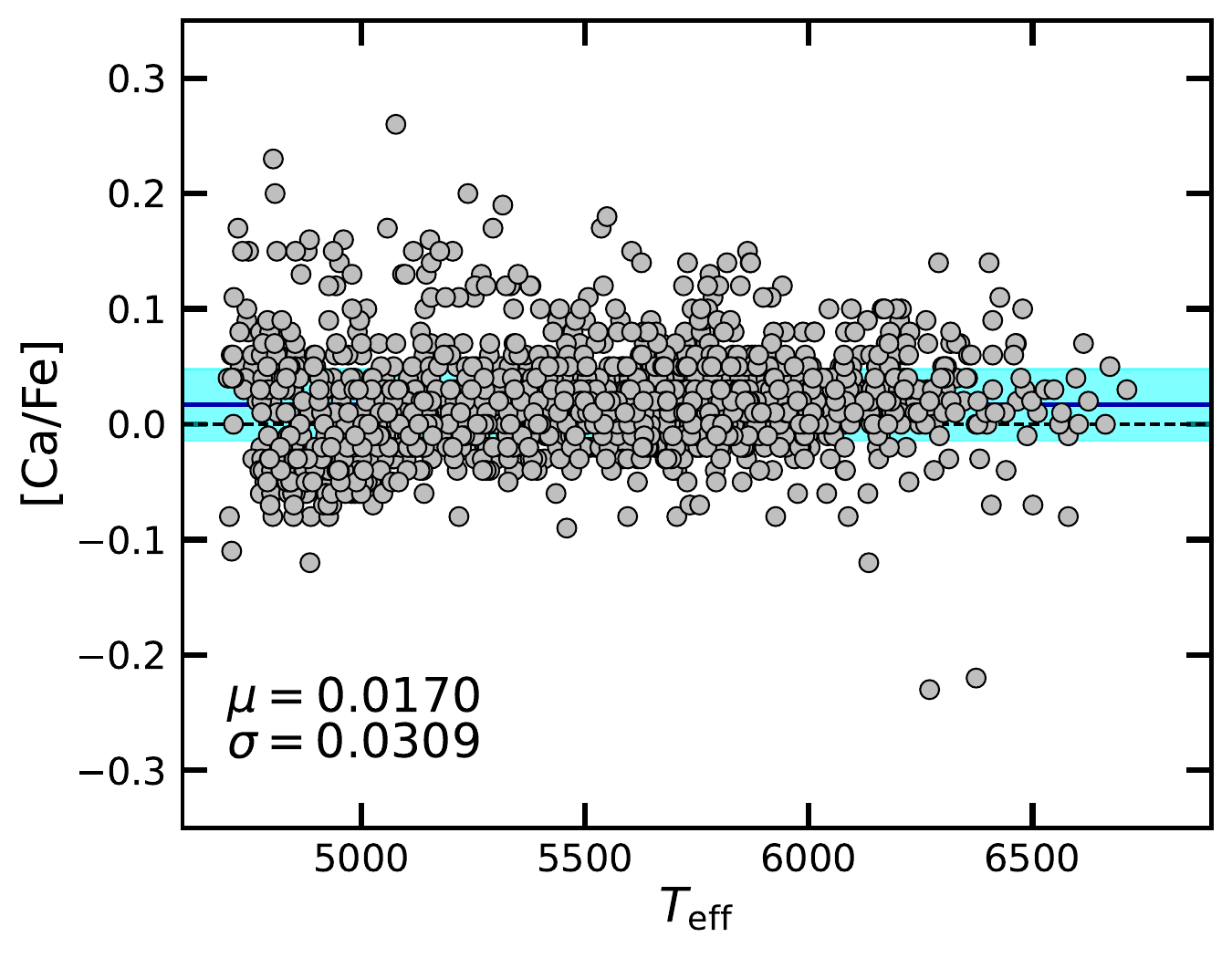}  
\caption{Abundance of calcium relative to iron as a function of effective temperature for the \cite{Brewer2016} sample.}
    \label{fig:brewer}
\end{figure}

We extended the analysis to a larger sample using the surface abundances determined by \cite{Brewer2016} for more than 1600 FGK stars. Figure \ref{fig:brewer} shows the calcium abundances of the \cite{Brewer2016} sample. Although we see that there is a tendency for calcium to be 0.02~dex higher than iron for the whole sample, there is no tendency for the abundance of calcium to increase with effective temperature. To confirm this, we performed an Orthogonal Distance Regression (ODR) statistical test. For our inferred data, we find that for stars with $\Teff\geq6000$~K, the fit gives a slope of $7.5\cdot10^{-5}$ with a correlation of ~0.35. For the \cite{Brewer2016} sample the slope is smaller ($7.5\cdot10^{-6}$) with almost no correlation (about ~0.02). 
This suggests that the transport processes competing with atomic diffusion are probably less efficient than what we included in grid C (turbulent mixing coefficient calibrated by VSA19 on helium abundances obtained from the analysis of seismic glitch signature) or much more efficient with a reference depth deeper than the region where radiative accelerations of Calcium are larger than gravity (see Fig.~\ref{fig:rad_profile2}). 
Another possibility could be that KIC~12317678 is a chemically peculiar star for which the competing transport processes are more efficient, leading to a strong impact of radiative accelerations on calcium. 

\section{Conclusion}
\label{sec:conclusion}

The purpose of this study was to explore the effect of radiative acceleration on the characterization of FGK-type stars. As mentioned, the standard method for computing radiative accelerations in MESA is very computationally expensive and not practical for the construction of a large number of models, which are necessary for stellar characterization procedures. To improve the calculation, we implemented the SVP method, which enables faster calculations (20 times faster compared to the standard method). This allowed us to include the effects of radiative accelerations for 12 chemical elements, and we show that they follow a similar evolution compared to the standard method.

We computed three grids. The first (grid A) neglects atomic diffusion for models of F-type stars. The second (grid B) considered $\Dtfe$, and the last (grid C) considered the SVP method. Compared to the standard method (grid A), the results obtained with the model that includes the SVP method (grid C) are consistent with those found in Paper II when we tested the results with $\Dtfe$ (comparison of grid A with grid B). The results show a bias of less than 1\% and a dispersion of 2.6\%, 0.9\%, and 7.2\% for the mass, radius, and age, respectively. In this case, we still find an increase in dispersion with stellar mass, confirming that the uncertainties for F-type stars are due to the neglect of atomic diffusion. This leads to uncertainties of up to 10\%, 4\%, and 29\% in the inference of mass, radius, and age, respectively, for individual stars.

Compared to the $\Dtfe$ results (grid B), the results obtained with the SVP method (grid C) show differences for higher-mass stars of less than 5\%, indicating that for F-type stars both methods do account for the effect of radiative accelerations and are equivalent for the inference of the fundamental properties. In contrast, there is a significant difference for lower-mass stars, where a larger dispersion leads to uncertainties of up to 8\%, 3\%, and 10\% in mass, radius, and age, respectively. This dispersion is not due to the inclusion of radiative accelerations, but rather to different turbulent mixing prescriptions for this mass regime. In the case of the SVP method, the turbulent mixing was calibrated to reproduce the lithium abundance at the surface of the Sun. This shows that neglecting the constraints brought by lithium on transport can lead to non-negligible uncertainties in the inference of stellar parameters.

Neither method ($\Dtfe$ and SVP) shows any major discrepancies in the prediction of surface abundances, except for calcium. In F-type stars, the combination of the SVP method with turbulent mixing led to an accumulation or constant abundance of this element at the stellar surface, while with the $\Dtfe$ method it led to a depletion. The models that use the SVP method show better agreement~with the \cite{morel2021} results. For KIC 12317678 specifically, this method yields a calcium abundance closer to the observed value. This suggests, as expected, that the SVP method provides a more realistic prediction of the evolution of the surface abundance compared to those that  use $\Dtfe$. Comparing the predictions of the models the include the SVP method and the observations, there is a higher abundance for the F-type stars compared to the low effective temperature stars. However, \cite{morel2021} studied a sample with a low number of F-type stars, which limits our ability to draw statistical conclusions. When we examined the variation in calcium abundance with effective temperature in a larger sample from \cite{Brewer2016}, we found no evidence that calcium accumulates in F-type stars. This suggests that the turbulent mixing used in grid C may not be suitable. 

The results of this work show that radiative accelerations need to be included in stellar models in order to accurately characterize F-type stars. The SVP method we used is fast and gives good results, on par  with those found using the current method implemented in MESA. Moreover, the implementation of radiative accelerations in stellar modes will be necessary to study the hotter stars that will be observed by upcoming missions such as Ariel and PLATO. 

\begin{acknowledgements}
This work was supported by Fundação para a Ciência e a Tecnologia (FCT) through the research grants UIDB/04434/2020 and UIDP/04434/2020. DOI: 10.54499/UIDB/04434/2020 and DOI: 10.54499/UIDP/04434/2020. NM acknowledges support from the Fundação para a Ciência e a Tecnologia (FCT) through the Fellowship UI/BD/152075/2021 and POCH/FSE (EC). DB acknowledges funding support by the Italian Ministerial Grant PRIN 2022, “Radiative opacities for astrophysical applications”, no. 2022NEXMP8, CUP C53D23001220006. MD acknowledges the support of CNES, focused on PLATO. We thank G. Alecian and F. LeBlanc for making SVP routines publicly available. We thank the anonymous referee for the valuable comments which helped to improve the paper.

\end{acknowledgements}

\bibliographystyle{aa} 
\bibliography{references} 

\begin{thebibliography}{52}
\expandafter\ifx\csname natexlab\endcsname\relax\def\natexlab#1{#1}\fi

\bibitem[{{Ahumada} {et~al.}(2020){Ahumada}, {Prieto}, {Almeida}, {Anders},
  {Anderson}, {Andrews}, {Anguiano}, {Arcodia}, {Armengaud}, {Aubert}, {Avila},
  {Avila-Reese}, {Badenes}, {Balland}, {Barger}, {Barrera-Ballesteros}, {Basu},
  {Bautista}, {Beaton}, {Beers}, {Benavides}, {Bender}, {Bernardi}, {Bershady},
  {Beutler}, {Bidin}, {Bird}, {Bizyaev}, {Blanc}, {Blanton}, {Boquien},
  {Borissova}, {Bovy}, {Brandt}, {Brinkmann}, {Brownstein}, {Bundy}, {Bureau},
  {Burgasser}, {Burtin}, {Cano-D{\'\i}az}, {Capasso}, {Cappellari}, {Carrera},
  {Chabanier}, {Chaplin}, {Chapman}, {Cherinka}, {Chiappini}, {Doohyun Choi},
  {Chojnowski}, {Chung}, {Clerc}, {Coffey}, {Comerford}, {Comparat}, {da
  Costa}, {Cousinou}, {Covey}, {Crane}, {Cunha}, {Ilha}, {Dai}, {Damsted},
  {Darling}, {Davidson}, {Davies}, {Dawson}, {De}, {de la Macorra}, {De Lee},
  {Queiroz}, {Deconto Machado}, {de la Torre}, {Dell'Agli}, {du Mas des
  Bourboux}, {Diamond-Stanic}, {Dillon}, {Donor}, {Drory}, {Duckworth},
  {Dwelly}, {Ebelke}, {Eftekharzadeh}, {Davis Eigenbrot}, {Elsworth},
  {Eracleous}, {Erfanianfar}, {Escoffier}, {Fan}, {Farr},
  {Fern{\'a}ndez-Trincado}, {Feuillet}, {Finoguenov}, {Fofie},
  {Fraser-McKelvie}, {Frinchaboy}, {Fromenteau}, {Fu}, {Galbany}, {Garcia},
  {Garc{\'\i}a-Hern{\'a}ndez}, {Oehmichen}, {Ge}, {Maia}, {Geisler}, {Gelfand},
  {Goddy}, {Gonzalez-Perez}, {Grabowski}, {Green}, {Grier}, {Guo}, {Guy},
  {Harding}, {Hasselquist}, {Hawken}, {Hayes}, {Hearty}, {Hekker}, {Hogg},
  {Holtzman}, {Horta}, {Hou}, {Hsieh}, {Huber}, {Hunt}, {Chitham}, {Imig},
  {Jaber}, {Angel}, {Johnson}, {Jones}, {J{\"o}nsson}, {Jullo}, {Kim},
  {Kinemuchi}, {Kirkpatrick}, {Kite}, {Klaene}, {Kneib}, {Kollmeier}, {Kong},
  {Kounkel}, {Krishnarao}, {Lacerna}, {Lan}, {Lane}, {Law}, {Le Goff}, {Leung},
  {Lewis}, {Li}, {Lian}, {Lin}, {Long}, {Longa-Pe{\~n}a}, {Lundgren}, {Lyke},
  {Ted Mackereth}, {MacLeod}, {Majewski}, {Manchado}, {Maraston}, {Martini},
  {Masseron}, {Masters}, {Mathur}, {McDermid}, {Merloni}, {Merrifield},
  {M{\'e}sz{\'a}ros}, {Miglio}, {Minniti}, {Minsley}, {Miyaji}, {Mohammad},
  {Mosser}, {Mueller}, {Muna}, {Mu{\~n}oz-Guti{\'e}rrez}, {Myers}, {Nadathur},
  {Nair}, {Nandra}, {do Nascimento}, {Nevin}, {Newman}, {Nidever}, {Nitschelm},
  {Noterdaeme}, {O'Connell}, {Olmstead}, {Oravetz}, {Oravetz}, {Osorio},
  {Pace}, {Padilla}, {Palanque-Delabrouille}, {Palicio}, {Pan}, {Pan},
  {Parker}, {Paviot}, {Peirani}, {Ram{\'r}ez}, {Penny}, {Percival},
  {Perez-Fournon}, {P{\'e}rez-R{\`a}fols}, {Petitjean}, {Pieri},
  {Pinsonneault}, {Poovelil}, {Povick}, {Prakash}, {Price-Whelan}, {Raddick},
  {Raichoor}, {Ray}, {Rembold}, {Rezaie}, {Riffel}, {Riffel}, {Rix}, {Robin},
  {Roman-Lopes}, {Rom{\'a}n-Z{\'u}{\~n}iga}, {Rose}, {Ross}, {Rossi},
  {Rowlands}, {Rubin}, {Salvato}, {S{\'a}nchez}, {S{\'a}nchez-Menguiano},
  {S{\'a}nchez-Gallego}, {Sayres}, {Schaefer}, {Schiavon}, {Schimoia},
  {Schlafly}, {Schlegel}, {Schneider}, {Schultheis}, {Schwope}, {Seo},
  {Serenelli}, {Shafieloo}, {Shamsi}, {Shao}, {Shen}, {Shetrone}, {Shirley},
  {Aguirre}, {Simon}, {Skrutskie}, {Slosar}, {Smethurst}, {Sobeck}, {Sodi},
  {Souto}, {Stark}, {Stassun}, {Steinmetz}, {Stello}, {Stermer},
  {Storchi-Bergmann}, {Streblyanska}, {Stringfellow}, {Stutz}, {Su{\'a}rez},
  {Sun}, {Taghizadeh-Popp}, {Talbot}, {Tayar}, {Thakar}, {Theriault}, {Thomas},
  {Thomas}, {Tinker}, {Tojeiro}, {Toledo}, {Tremonti}, {Troup}, {Tuttle},
  {Unda-Sanzana}, {Valentini}, {Vargas-Gonz{\'a}lez}, {Vargas-Maga{\~n}a},
  {V{\'a}zquez-Mata}, {Vivek}, {Wake}, {Wang}, {Weaver}, {Weijmans}, {Wild},
  {Wilson}, {Wilson}, {Wolthuis}, {Wood-Vasey}, {Yan}, {Yang}, {Y{\`e}che},
  {Zamora}, {Zarrouk}, {Zasowski}, {Zhang}, {Zhao}, {Zhao}, {Zheng}, {Zheng},
  {Zhu}, \& {Zou}}]{Ahumada2020}
{Ahumada}, R., {Prieto}, C.~A., {Almeida}, A., {et~al.} 2020, \apjs, 249, 3

\bibitem[{{Alecian} \& {LeBlanc}(2020)}]{alecian20}
{Alecian}, G. \& {LeBlanc}, F. 2020, \mnras, 498, 3420

\bibitem[{{Asplund} {et~al.}(2009){Asplund}, {Grevesse}, {Sauval}, \&
  {Scott}}]{Asplund2009}
{Asplund}, M., {Grevesse}, N., {Sauval}, A.~J., \& {Scott}, P. 2009, \araa, 47,
  481

\bibitem[{{Ball} \& {Gizon}(2014)}]{Ball2014}
{Ball}, W.~H. \& {Gizon}, L. 2014, \aap, 568, A123

\bibitem[{{Borucki} {et~al.}(2010){Borucki}, {Koch}, \& et~al.}]{kepler}
{Borucki}, W.~J., {Koch}, D., \& et~al. 2010, Science, 327, 977

\bibitem[{{Brewer} {et~al.}(2016){Brewer}, {Fischer}, {Valenti}, \&
  {Piskunov}}]{Brewer2016}
{Brewer}, J.~M., {Fischer}, D.~A., {Valenti}, J.~A., \& {Piskunov}, N. 2016,
  \apjs, 225, 32

\bibitem[{{Campilho} {et~al.}(2022){Campilho}, {Deal}, \&
  {Bossini}}]{campilho22}
{Campilho}, B., {Deal}, M., \& {Bossini}, D. 2022, \aap, 659, A162

\bibitem[{{Chaboyer} {et~al.}(2001){Chaboyer}, {Fenton}, {Nelan}, {Patnaude},
  \& {Simon}}]{Chaboyer2001}
{Chaboyer}, B., {Fenton}, W.~H., {Nelan}, J.~E., {Patnaude}, D.~J., \& {Simon},
  F.~E. 2001, \apj, 562, 521

\bibitem[{{Cox} \& {Giuli}(1968)}]{Cox1968}
{Cox}, J.~P. \& {Giuli}, R.~T. 1968, {Principles of stellar structure} (Gordon
  \& Breach)

\bibitem[{{Cunha} {et~al.}(2021){Cunha}, {Roxburgh}, {Aguirre B{\o}rsen-Koch},
  {Ball}, {Basu}, {Chaplin}, {Goupil}, {Nsamba}, {Ong}, {Reese}, {Verma},
  {Belkacem}, {Campante}, {Christensen-Dalsgaard}, {Clara}, {Deheuvels},
  {Monteiro}, {Noll}, {Ouazzani}, {R{\o}rsted}, {Stokholm}, \&
  {Winther}}]{Cunha2021}
{Cunha}, M.~S., {Roxburgh}, I.~W., {Aguirre B{\o}rsen-Koch}, V., {et~al.} 2021,
  \mnras, 508, 5864

\bibitem[{{Cupani} {et~al.}(2017){Cupani}, {D'Odorico}, {Cristiani},
  {Gonz{\'a}lez-Hern{\'a}ndez}, {Lovis}, {Sousa}, {Di Marcantonio}, \&
  {M{\'e}gevand}}]{ESPRESSO}
{Cupani}, G., {D'Odorico}, V., {Cristiani}, S., {et~al.} 2017, in Astronomical
  Society of the Pacific Conference Series, Vol. 512, Astronomical Data
  Analysis Software and Systems XXV, ed. N.~P.~F. {Lorente}, K.~{Shortridge},
  \& R.~{Wayth}, 209

\bibitem[{{Davies} {et~al.}(2016){Davies}, {Silva Aguirre}, {Bedding},
  {Handberg}, {Lund}, {Chaplin}, {Huber}, {White}, {Benomar}, {Hekker}, {Basu},
  {Campante}, {Christensen-Dalsgaard}, {Elsworth}, {Karoff}, {Kjeldsen},
  {Lundkvist}, {Metcalfe}, \& {Stello}}]{Davies2016}
{Davies}, G.~R., {Silva Aguirre}, V., {Bedding}, T.~R., {et~al.} 2016, \mnras,
  456, 2183

\bibitem[{{Deal} {et~al.}(2015){Deal}, {Richard}, \& {Vauclair}}]{Deal2015}
{Deal}, M., {Richard}, O., \& {Vauclair}, S. 2015, \aap, 584, A105

\bibitem[{{Ferguson} {et~al.}(2005){Ferguson}, {Alexander}, {Allard}, {Barman},
  {Bodnarik}, {Hauschildt}, {Heffner-Wong}, \& {Tamanai}}]{Ferguson2005}
{Ferguson}, J.~W., {Alexander}, D.~R., {Allard}, F., {et~al.} 2005, \apj, 623,
  585

\bibitem[{{Gaia Collaboration} {et~al.}(2021){Gaia Collaboration}, {Brown},
  {Vallenari}, {Prusti}, {de Bruijne}, {Babusiaux}, {Biermann}, {Creevey},
  {Evans}, {Eyer}, {Hutton}, {Jansen}, {Jordi}, {Klioner}, {Lammers},
  {Lindegren}, {Luri}, {Mignard}, {Panem}, {Pourbaix}, {Randich}, {Sartoretti},
  {Soubiran}, {Walton}, {Arenou}, {Bailer-Jones}, {Bastian}, {Cropper},
  {Drimmel}, {Katz}, {Lattanzi}, {van Leeuwen}, {Bakker}, {Cacciari},
  {Casta{\~n}eda}, {De Angeli}, {Ducourant}, {Fabricius}, {Fouesneau},
  {Fr{\'e}mat}, {Guerra}, {Guerrier}, {Guiraud}, {Jean-Antoine Piccolo},
  {Masana}, {Messineo}, {Mowlavi}, {Nicolas}, {Nienartowicz}, {Pailler},
  {Panuzzo}, {Riclet}, {Roux}, {Seabroke}, {Sordo}, {Tanga}, {Th{\'e}venin},
  {Gracia-Abril}, {Portell}, {Teyssier}, {Altmann}, {Andrae}, {Bellas-Velidis},
  {Benson}, {Berthier}, {Blomme}, {Brugaletta}, {Burgess}, {Busso}, {Carry},
  {Cellino}, {Cheek}, {Clementini}, {Damerdji}, {Davidson}, {Delchambre},
  {Dell'Oro}, {Fern{\'a}ndez-Hern{\'a}ndez}, {Galluccio}, {Garc{\'\i}a-Lario},
  {Garcia-Reinaldos}, {Gonz{\'a}lez-N{\'u}{\~n}ez}, {Gosset}, {Haigron},
  {Halbwachs}, {Hambly}, {Harrison}, {Hatzidimitriou}, {Heiter},
  {Hern{\'a}ndez}, {Hestroffer}, {Hodgkin}, {Holl}, {Jan{\ss}en}, {Jevardat de
  Fombelle}, {Jordan}, {Krone-Martins}, {Lanzafame}, {L{\"o}ffler}, {Lorca},
  {Manteiga}, {Marchal}, {Marrese}, {Moitinho}, {Mora}, {Muinonen}, {Osborne},
  {Pancino}, {Pauwels}, {Petit}, {Recio-Blanco}, {Richards}, {Riello},
  {Rimoldini}, {Robin}, {Roegiers}, {Rybizki}, {Sarro}, {Siopis}, {Smith},
  {Sozzetti}, {Ulla}, {Utrilla}, {van Leeuwen}, {van Reeven}, {Abbas}, {Abreu
  Aramburu}, {Accart}, {Aerts}, {Aguado}, {Ajaj}, {Altavilla}, {{\'A}lvarez},
  {{\'A}lvarez Cid-Fuentes}, {Alves}, {Anderson}, {Anglada Varela}, {Antoja},
  {Audard}, {Baines}, {Baker}, {Balaguer-N{\'u}{\~n}ez}, {Balbinot}, {Balog},
  {Barache}, {Barbato}, {Barros}, {Barstow}, {Bartolom{\'e}}, {Bassilana},
  {Bauchet}, {Baudesson-Stella}, {Becciani}, {Bellazzini}, {Bernet}, {Bertone},
  {Bianchi}, {Blanco-Cuaresma}, {Boch}, {Bombrun}, {Bossini}, {Bouquillon},
  {Bragaglia}, {Bramante}, {Breedt}, {Bressan}, {Brouillet}, {Bucciarelli},
  {Burlacu}, {Busonero}, {Butkevich}, {Buzzi}, {Caffau}, {Cancelliere},
  {C{\'a}novas}, {Cantat-Gaudin}, {Carballo}, {Carlucci}, {Carnerero},
  {Carrasco}, {Casamiquela}, {Castellani}, {Castro-Ginard}, {Castro Sampol},
  {Chaoul}, {Charlot}, {Chemin}, {Chiavassa}, {Cioni}, {Comoretto}, {Cooper},
  {Cornez}, {Cowell}, {Crifo}, {Crosta}, {Crowley}, {Dafonte}, {Dapergolas},
  {David}, {David}, {de Laverny}, {De Luise}, {De March}, {De Ridder}, {de
  Souza}, {de Teodoro}, {de Torres}, {del Peloso}, {del Pozo}, {Delbo},
  {Delgado}, {Delgado}, {Delisle}, {Di Matteo}, {Diakite}, {Diener},
  {Distefano}, {Dolding}, {Eappachen}, {Edvardsson}, {Enke}, {Esquej}, {Fabre},
  {Fabrizio}, {Faigler}, {Fedorets}, {Fernique}, {Fienga}, {Figueras},
  {Fouron}, {Fragkoudi}, {Fraile}, {Franke}, {Gai}, {Garabato},
  {Garcia-Gutierrez}, {Garc{\'\i}a-Torres}, {Garofalo}, {Gavras}, {Gerlach},
  {Geyer}, {Giacobbe}, {Gilmore}, {Girona}, {Giuffrida}, {Gomel}, {Gomez},
  {Gonzalez-Santamaria}, {Gonz{\'a}lez-Vidal}, {Granvik},
  {Guti{\'e}rrez-S{\'a}nchez}, {Guy}, {Hauser}, {Haywood}, {Helmi}, {Hidalgo},
  {Hilger}, {H{\l}adczuk}, {Hobbs}, {Holland}, {Huckle}, {Jasniewicz},
  {Jonker}, {Juaristi Campillo}, {Julbe}, {Karbevska}, {Kervella}, {Khanna},
  {Kochoska}, {Kontizas}, {Kordopatis}, {Korn}, {Kostrzewa-Rutkowska},
  {Kruszy{\'n}ska}, {Lambert}, {Lanza}, {Lasne}, {Le Campion}, {Le Fustec},
  {Lebreton}, {Lebzelter}, {Leccia}, {Leclerc}, {Lecoeur-Taibi}, {Liao},
  {Licata}, {Lindstr{\o}m}, {Lister}, {Livanou}, {Lobel}, {Madrero Pardo},
  {Managau}, {Mann}, {Marchant}, {Marconi}, {Marcos Santos}, {Marinoni},
  {Marocco}, {Marshall}, {Martin Polo}, {Mart{\'\i}n-Fleitas}, {Masip},
  {Massari}, {Mastrobuono-Battisti}, {Mazeh}, {McMillan}, {Messina},
  {Michalik}, {Millar}, {Mints}, {Molina}, {Molinaro}, {Moln{\'a}r},
  {Montegriffo}, {Mor}, {Morbidelli}, {Morel}, {Morris}, {Mulone}, {Munoz},
  {Muraveva}, {Murphy}, {Musella}, {Noval}, {Ord{\'e}novic}, {Orr{\`u}},
  {Osinde}, {Pagani}, {Pagano}, {Palaversa}, {Palicio}, {Panahi}, {Pawlak},
  {Pe{\~n}alosa Esteller}, {Penttil{\"a}}, {Piersimoni}, {Pineau}, {Plachy},
  {Plum}, {Poggio}, {Poretti}, {Poujoulet}, {Pr{\v{s}}a}, {Pulone}, {Racero},
  {Ragaini}, {Rainer}, {Raiteri}, {Rambaux}, {Ramos}, {Ramos-Lerate}, {Re
  Fiorentin}, {Regibo}, {Reyl{\'e}}, {Ripepi}, {Riva}, {Rixon}, {Robichon},
  {Robin}, {Roelens}, {Rohrbasser}, {Romero-G{\'o}mez}, {Rowell}, {Royer},
  {Rybicki}, {Sadowski}, {Sagrist{\`a} Sell{\'e}s}, {Sahlmann}, {Salgado},
  {Salguero}, {Samaras}, {Sanchez Gimenez}, {Sanna}, {Santove{\~n}a},
  {Sarasso}, {Schultheis}, {Sciacca}, {Segol}, {Segovia}, {S{\'e}gransan},
  {Semeux}, {Shahaf}, {Siddiqui}, {Siebert}, {Siltala}, {Slezak}, {Smart},
  {Solano}, {Solitro}, {Souami}, {Souchay}, {Spagna}, {Spoto}, {Steele},
  {Steidelm{\"u}ller}, {Stephenson}, {S{\"u}veges}, {Szabados}, {Szegedi-Elek},
  {Taris}, {Tauran}, {Taylor}, {Teixeira}, {Thuillot}, {Tonello}, {Torra},
  {Torra}, {Turon}, {Unger}, {Vaillant}, {van Dillen}, {Vanel}, {Vecchiato},
  {Viala}, {Vicente}, {Voutsinas}, {Weiler}, {Wevers}, {Wyrzykowski}, {Yoldas},
  {Yvard}, {Zhao}, {Zorec}, {Zucker}, {Zurbach}, \&
  {Zwitter}}]{Gaia_Collaboration2021}
{Gaia Collaboration}, {Brown}, A.~G.~A., {Vallenari}, A., {et~al.} 2021, \aap,
  649, A1

\bibitem[{{Herwig}(2000)}]{Herwig2000}
{Herwig}, F. 2000, \aap, 360, 952

\bibitem[{{Hu} {et~al.}(2011){Hu}, {Tout}, {Glebbeek}, \& {Dupret}}]{Hu2011}
{Hu}, H., {Tout}, C.~A., {Glebbeek}, E., \& {Dupret}, M.~A. 2011, \mnras, 418,
  195

\bibitem[{{Hui-Bon-Hoa}(2024)}]{huibonhoa24}
{Hui-Bon-Hoa}, A. 2024, \aap, 691, A266

\bibitem[{{Iglesias} \& {Rogers}(1996)}]{Iglesias1996}
{Iglesias}, C.~A. \& {Rogers}, F.~J. 1996, \apj, 464, 943

\bibitem[{{Jermyn} {et~al.}(2023){Jermyn}, {Bauer}, {Schwab}, {Farmer}, {Ball},
  {Bellinger}, {Dotter}, {Joyce}, {Marchant}, {Mombarg}, {Wolf}, {Sunny Wong},
  {Cinquegrana}, {Farrell}, {Smolec}, {Thoul}, {Cantiello}, {Herwig}, {Toloza},
  {Bildsten}, {Townsend}, \& {Timmes}}]{Jermyn2023}
{Jermyn}, A.~S., {Bauer}, E.~B., {Schwab}, J., {et~al.} 2023, \apjs, 265, 15

\bibitem[{{Krishna Swamy}(1966)}]{Krishna1966}
{Krishna Swamy}, K.~S. 1966, \apj, 145, 174

\bibitem[{{LeBlanc} \& {Alecian}(2004)}]{LeBlanc2004}
{LeBlanc}, F. \& {Alecian}, G. 2004, \mnras, 352, 1329

\bibitem[{{Lund} {et~al.}(2017){Lund}, {Silva Aguirre}, {Davies}, {Chaplin},
  {Christensen-Dalsgaard}, {Houdek}, {White}, {Bedding}, {Ball}, {Huber},
  {Antia}, {Lebreton}, {Latham}, {Handberg}, {Verma}, {Basu}, {Casagrande},
  {Justesen}, {Kjeldsen}, \& {Mosumgaard}}]{Lund2017}
{Lund}, M.~N., {Silva Aguirre}, V., {Davies}, G.~R., {et~al.} 2017, \apj, 835,
  172

\bibitem[{{Michaud} {et~al.}(2015){Michaud}, {Alecian}, \&
  {Richer}}]{michaud15}
{Michaud}, G., {Alecian}, G., \& {Richer}, J. 2015, {Atomic Diffusion in Stars}
  (Springer)

\bibitem[{{Michaud} {et~al.}(2011{\natexlab{a}}){Michaud}, {Richer}, \&
  {Richard}}]{Michaud2011a}
{Michaud}, G., {Richer}, J., \& {Richard}, O. 2011{\natexlab{a}}, \aap, 529,
  A60

\bibitem[{{Michaud} {et~al.}(2011{\natexlab{b}}){Michaud}, {Richer}, \&
  {Vick}}]{Michaud2011b}
{Michaud}, G., {Richer}, J., \& {Vick}, M. 2011{\natexlab{b}}, \aap, 534, A18

\bibitem[{{Moedas} {et~al.}(2024){Moedas}, {Bossini}, {Deal}, \&
  {Cunha}}]{Moedas2024}
{Moedas}, N., {Bossini}, D., {Deal}, M., \& {Cunha}, M.~S. 2024, \aap, 684,
  A113

\bibitem[{{Moedas} {et~al.}(2022){Moedas}, {Deal}, {Bossini}, \&
  {Campilho}}]{moedas2022}
{Moedas}, N., {Deal}, M., {Bossini}, D., \& {Campilho}, B. 2022, \aap, 666, A43

\bibitem[{{Mombarg} {et~al.}(2022){Mombarg}, {Dotter}, {Rieutord},
  {Michielsen}, {Van Reeth}, \& {Aerts}}]{Mombarg2022}
{Mombarg}, J. S.~G., {Dotter}, A., {Rieutord}, M., {et~al.} 2022, \apj, 925,
  154

\bibitem[{{Morel} {et~al.}(2021){Morel}, {Creevey}, {Montalb{\'a}n}, {Miglio},
  \& {Willett}}]{morel2021}
{Morel}, T., {Creevey}, O.~L., {Montalb{\'a}n}, J., {Miglio}, A., \& {Willett},
  E. 2021, \aap, 646, A78

\bibitem[{{Nsamba} {et~al.}(2018){Nsamba}, {Campante}, {Monteiro}, {Cunha},
  {Rendle}, {Reese}, \& {Verma}}]{Nsamba2018}
{Nsamba}, B., {Campante}, T.~L., {Monteiro}, M.~J.~P.~F.~G., {et~al.} 2018,
  \mnras, 477, 5052

\bibitem[{{Paxton} {et~al.}(2011){Paxton}, {Bildsten}, {Dotter}, \&
  et~al.}]{Paxton2011}
{Paxton}, B., {Bildsten}, L., {Dotter}, A., \& et~al. 2011, \apjs, 192, 3

\bibitem[{{Paxton} {et~al.}(2013){Paxton}, {Cantiello}, {Arras}, {Bildsten},
  {Brown}, {Dotter}, {Mankovich}, {Montgomery}, {Stello}, {Timmes}, \&
  {Townsend}}]{Paxton2013}
{Paxton}, B., {Cantiello}, M., {Arras}, P., {et~al.} 2013, \apjs, 208, 4

\bibitem[{{Paxton} {et~al.}(2015){Paxton}, {Marchant}, {Schwab}, {Bauer},
  {Bildsten}, {Cantiello}, {Dessart}, {Farmer}, {Hu}, {Langer}, {Townsend},
  {Townsley}, \& {Timmes}}]{Paxton2015}
{Paxton}, B., {Marchant}, P., {Schwab}, J., {et~al.} 2015, \apjs, 220, 15

\bibitem[{{Paxton} {et~al.}(2018){Paxton}, {Schwab}, {Bauer}, {Bildsten},
  {Blinnikov}, {Duffell}, {Farmer}, {Goldberg}, {Marchant}, {Sorokina},
  {Thoul}, {Townsend}, \& {Timmes}}]{Paxton2018}
{Paxton}, B., {Schwab}, J., {Bauer}, E.~B., {et~al.} 2018, \apjs, 234, 34

\bibitem[{{Paxton} {et~al.}(2019){Paxton}, {Smolec}, {Schwab}, {Gautschy},
  {Bildsten}, {Cantiello}, {Dotter}, {Farmer}, {Goldberg}, {Jermyn}, {Kanbur},
  {Marchant}, {Thoul}, {Townsend}, {Wolf}, {Zhang}, \& {Timmes}}]{Paxton2019}
{Paxton}, B., {Smolec}, R., {Schwab}, J., {et~al.} 2019, \apjs, 243, 10

\bibitem[{{Proffitt} \& {Michaud}(1991)}]{Proffitt1991}
{Proffitt}, C.~R. \& {Michaud}, G. 1991, \apj, 371, 584

\bibitem[{{Rauer} {et~al.}(2024){Rauer}, {Aerts}, {Cabrera}, {Deleuil},
  {Erikson}, {Gizon}, {Goupil}, {Heras}, {Lorenzo-Alvarez}, {Marliani},
  {Martin-Garcia}, {Mas-Hesse}, {O'Rourke}, {Osborn}, {Pagano}, {Piotto},
  {Pollacco}, {Ragazzoni}, {Ramsay}, {Udry}, {Appourchaux}, {Benz},
  {Brandeker}, {G{\"u}del}, {Janot-Pacheco}, {Kabath}, {Kjeldsen}, {Min},
  {Santos}, {Smith}, {Suarez}, {Werner}, {Aboudan}, {Abreu}, {Acu{\~n}a},
  {Adams}, {Adibekyan}, {Affer}, {Agneray}, {Agnor}, {Aguirre B{\o}rsen-Koch},
  {Ahmed}, {Aigrain}, {Al-Bahlawan}, {Alcacera Gil}, {Alei}, {Alencar},
  {Alexander}, {Alfonso-Garz{\'o}n}, {Alibert}, {Allende Prieto}, {Almeida},
  {Alonso Sobrino}, {Altavilla}, {Althaus}, {Alonso Alvarez Trujillo},
  {Amarsi}, {Ammler-von Eiff}, {Am{\^o}res}, {Andrade}, {Antoniadis-Karnavas},
  {Ant{\'o}nio}, {Aparicio del Moral}, {Appolloni}, {Arena}, {Armstrong},
  {Aroca Aliaga}, {Asplund}, {Audenaert}, {Auricchio}, {Avelino}, {Baeke},
  {Bailli{\'e}}, {Balado}, {Balestra}, {Ball}, {Ballans}, {Ballot}, {Barban},
  {Barbary}, {Barbieri}, {Barcel{\'o} Forteza}, {Barker}, {Barklem}, {Barnes},
  {Barrado Navascues}, {Barragan}, {Baruteau}, {Basu}, {Baudin}, {Baumeister},
  {Bayliss}, {Bazot}, {Beck}, {Bedding}, {Belkacem}, {Bellinger}, {Benatti},
  {Benomar}, {B{\'e}rard}, {Bergemann}, {Bergomi}, {Bernardo}, {Biazzo},
  {Bignamini}, {Bigot}, {Billot}, {Binet}, {Biondi}, {Biondi}, {Birch},
  {Bitsch}, {Bluhm Ceballos}, {B{\'o}di}, {Bogn{\'a}r}, {Boisse}, {Bolmont},
  {Bonanno}, {Bonavita}, {Bonfanti}, {Bonfils}, {Bonito}, {Bonomo},
  {B{\"o}rner}, {Boro Saikia}, {Borreguero Mart{\'\i}n}, {Borsa}, {Borsato},
  {Bossini}, {Bouchy}, {Bou{\'e}}, {Boufleur}, {Boumier}, {Bourrier}, {Bowman},
  {Bozzo}, {Bradley}, {Bray}, {Bressan}, {Breton}, {Brienza}, {Brito}, {Brogi},
  {Brown}, {Brown}, {Brun}, {Bruno}, {Bruns}, {Buchhave}, {Bugnet}, {Buldgen},
  {Burgess}, {Busatta}, {Busso}, {Buzasi}, {Caballero}, {Cabral}, {Calderone},
  {Cameron}, {Cameron}, {Campante}, {Canto Martins}, {Cara}, {Carone},
  {Carrasco}, {Casagrande}, {Casewell}, {Cassisi}, {Castellani}, {Castro},
  {Catala}, {Catal{\'a}n Fern{\'a}ndez}, {Catelan}, {Cegla}, {Cerruti},
  {Cessa}, {Chadid}, {Chaplin}, {Charpinet}, {Chiappini}, {Chiarucci},
  {Chiavassa}, {Chinellato}, {Chirulli}, {Christensen-Dalsgaard}, {Church},
  {Claret}, {Clarke}, {Claudi}, {Clermont}, {Coelho}, {Coelho}, {Cogato},
  {Colom{\'e}}, {Condamin}, {Conseil}, {Corbard}, {Correia}, {Corsaro},
  {Cosentino}, {Costes}, {Cottinelli}, {Covone}, {Creevey}, {Crida},
  {Csizmadia}, {Cunha}, {Curry}, {da Costa}, {da Silva}, {Dalal}, {Damasso},
  {Damiani}, {Damiani}, {Liduina das Chagas}, {Davies}, {Davies}, {Davies},
  {Davison}, {de Almeida}, {de Angeli}, {Cabral de Barros}, {de Castro
  Le{\~a}o}, {Brito de Freitas}, {de Freitas}, {De Martino}, {Renan de
  Medeiros}, {de Paula}, {de Plaa}, {De Ridder}, {Deal}, {Decin}, {Deeg},
  {Degl'Innocenti}, {Deheuvels}, {del Burgo}, {Del Sordo}, {Delgado-Mena},
  {Demangeon}, {Denk}, {Derekas}, {Desidera}, {Dexet}, {Di Criscienzo}, {Di
  Giorgio}, {Di Mauro}, {Diaz Rial}, {D{\'\i}az-Garc{\'\i}a}, {Dima},
  {Dinuzzi}, {Dionatos}, {Distefano}, {do Nascimento}, {Domingo}, {D'Orazi},
  {Dorn}, {Doyle}, {Duarte}, {Ducellier}, {Dumaye}, {Dumusque}, {Dupret},
  {Eggenberger}, {Ehrenreich}, {Eigm{\"u}ller}, {Eising}, {Emilio}, {Eriksson},
  {Ermocida}, {Isidoro Escate Giribaldi}, {Eschen}, {Estrela}, {Evans},
  {Fabbian}, {Fabrizio}, {Faria}, {Farina}, {Farinato}, {Feliz}, {Feltzing},
  {Fenouillet}, {Ferrari}, {Ferraz-Mello}, {Fialho}, {Fienga}, {Figueira},
  {Fiori}, {Flaccomio}, {Focardi}, {Foley}, {Fontignie}, {Ford}, {Fornazier},
  {Forveille}, {Fossati}, {de Marca Franca}, {da Silva}, {Frasca}, {Fridlund},
  {Furlan}, {Gabler}, {Gaido}, {Gallagher}, {Galli}, {Garcia}, {Garc{\'\i}a
  Hern{\'a}ndez}, {Garcia Munoz}, {Garc{\'\i}a-V{\'a}zquez}, {Garrido Haba},
  {Gaulme}, {Gauthier}, {Gehan}, {Gent}, {Georgieva}, {Ghigo}, {Giana}, {Gill},
  {Girardi}, {Giuliatti Winter}, {Giusi}, {Gomes da Silva}, {G{\'o}mez Zazo},
  {Gomez-Lopez}, {Isai Gonz{\'a}lez Hern{\'a}ndez}, {Gonzalez Murillo},
  {Gorius}, {Gouel}, {Goulty}, {Granata}, {Grenfell}, {Grie{\ss}bach},
  {Grolleau}, {Grouffal}, {Grziwa}, {Guarcello}, {Gueguen}, {Guenther},
  {Guilhem}, {Guillerot}, {Guiot}, {Guterman}, {Guti{\'e}rrez},
  {Guti{\'e}rrez-Canales}, {Hagelberg}, {Haldemann}, {Hall}, {Handberg},
  {Harrison}, {Harrison}, {Hasiba}, {Haswell}, {Hatalova}, {Hatzes}, {Haywood},
  {H{\'e}brard}, {Heckes}, {Heiter}, {Hekker}, {Heller}, {Helling},
  {Helminiak}, {Hemsley}, {Heng}, {Hermans}, {Hermes}, {Hidalgo Torres},
  {Hinkel}, {Hobbs}, {Hodgkin}, {Hofmann}, {Hojjatpanah}, {Houdek}, {Huber},
  {Huesler}, {Hui-Bon-Hoa}, {Huygen}, {Huynh}, {Iro}, {Irwin}, {Irwin},
  {Izidoro}, {Jacquinod}, {Emborg Jannsen}, {Janson}, {Jeszenszky}, {Jiang},
  {Jos{\'e} Jimenez Mancebo}, {Jofre}, {Johansen}, {Johnston}, {Jones},
  {Kallinger}, {K{\'a}lm{\'a}n}, {Kanitz}, {Karjalainen}, {Karjalainen},
  {Karoff}, {Kawaler}, {Kawata}, {Keereman}, {Keiderling}, {Kennedy},
  {Kenworthy}, {Kerschbaum}, {Kidger}, {Kiefer}, {Kintziger}, {Kislyakova},
  {Kiss}, {Klagyivik}, {Klahr}, {Klevas}, {Kochukhov}, {K{\"o}hler}, {Kolb},
  {Koncz}, {Korth}, {Kostogryz}, {Kov{\'a}cs}, {Kov{\'a}cs}, {Kozhura},
  {Krivova}, {Ku{\v{c}}inskas}, {Kuhlemann}, {Kupka}, {Laauwen}, {Labiano},
  {Lagarde}, {Laget}, {Laky}, {Lam}, {Lambrechts}, {Lammer}, {Lanza},
  {Lanzafame}, {Lares Martiz}, {Laskar}, {Latter}, {Lavanant}, {Lawrenson},
  {Lazzoni}, {Lebre}, {Lebreton}, {Lecavelier des Etangs}, {Leinhardt},
  {Leleu}, {Lendl}, {Leto}, {Levillain}, {Libert}, {Lichtenberg}, {Ligi},
  {Lignieres}, {Lillo-Box}, {Linsky}, {Scige Liu}, {Loidolt}, {Longval},
  {Lopes}, {Lorenzani}, {Ludwig}, {Lund}, {Sloth Lundkvist}, {Luri},
  {Maceroni}, {Madden}, {Madhusudhan}, {Maggio}, {Magliano}, {Magrin}, {Mahy},
  {Maibaum}, {Malac-Allain}, {Malapert}, {Malavolta}, {Maldonado}, {Mamonova},
  {Manchon}, {Mann}, {Mantovan}, {Marafatto}, {Marconi}, {Mardling}, {Marigo},
  {Marinoni}, {Marques}, {Marques}, {Marrese}, {Marshall}, {Mart{\'\i}nez
  Perales}, {Mary}, {Marzari}, {Masana}, {Mascher}, {Mathis}, {Mathur},
  {Mattiuci Figueiredo}, {Maxted}, {Mazeh}, {Mazevet}, {Mazzei}, {McCormac},
  {McMillan}, {Menou}, {Merle}, {Meru}, {Mesa}, {Messina}, {M{\'e}sz{\'a}ros},
  {Meunier}, {Meunier}, {Micela}, {Michaelis}, {Michel}, {Michielsen},
  {Michtchenko}, {Miglio}, {Miguel}, {Milligan}, {Mirouh}, {Mitchel}, {Moedas},
  {Molendini}, {Moln{\'a}r}, {Mombarg}, {Montalban}, {Montalto}, {Monteiro},
  {Morales}, {Morales-Calderon}, {Morbidelli}, {Mordasini}, {Moreau}, {Morel},
  {Morello}, {Morin}, {Mortier}, {Mosser}, {Mourard}, {Mousis}, {Moutou},
  {Mowlavi}, {Moya}, {Muehlmann}, {Muirhead}, {Munari}, {Musella}, {Mustill},
  {Nardetto}, {Nardiello}, {Narita}, {Nascimbeni}, {Nash}, {Neiner}, {Nelson},
  {Nettelmann}, {Nicolini}, {Nielsen}, {Niemi}, {Noack}, {Noels-Grotsch},
  {Noll}, {Norazman}, {Norton}, {Nsamba}, {Ofir}, {Ogilvie}, {Olander},
  {Olivetto}, {Olofsson}, {Ong}, {Ortolani}, {Oshagh}, {Ottacher},
  {Ottensamer}, {Ouazzani}, {Paardekooper}, {Pace}, {Pajas}, {Palacios},
  {Palandri}, {Palle}, {Paproth}, {Parro}, {Parviainen}, {Granado},
  {Passegger}, {Pastor-Morales}, {P{\"a}tzold}, {Gade Pedersen}, {Pena
  Hidalgo}, {Pepe}, {Pereira}, {Persson}, {Pertenais}, {Peter}, {Petit},
  {Petit}, {Pezzuto}, {Pichierri}, {Pietrinferni}, {Pinheiro}, {Pinsonneault},
  {Plachy}, {Plasson}, {Plez}, {Poppenhaeger}, {Poretti}, {Portaluri},
  {Portell}, {Frederico Porto de Mello}, {Poyatos}, {Pozuelos}, {Prada Moroni},
  {Pricopi}, {Prisinzano}, {Quade}, {Quirrenbach160}, {Rabanal Reina6},
  {Rabello Soares}, {Raimondo}, {Rainer}, {Ram{\'o}n Rod{\'o}n},
  {Ram{\'o}n-Ballesta}, {Ramos Zapata}, {R{\"a}tz}, {Rauterberg}, {Redman},
  {Redmer}, {Reese}, {Regibo}, {Reiners}, {Reinhold}, {Renie}, {Ribas},
  {Ribeiro}, {Pereira Ricciardi}, {Rice}, {Richard}, {Riello}, {Rieutord},
  {Ripepi}, {Rixon}, {Rockstein}, {Rodr{\'\i}guez}, {Rodr{\'\i}guez D{\'\i}az},
  {Rodriguez Garcia}, {Rodriguez-Gomez}, {Roehlly}, {Roig}, {Rojas-Ayala},
  {Rolf}, {Lysgaard R{\o}rsted}, {Rosado}, {Rosotti}, {Roth}, {Roth},
  {Rousseau}, {Roxburgh}, {Roy}, {Royer}, {Ruane}, {Rufini Mastropasqua}, {Ruiz
  de Galarreta}, {Russi}, {Saar}, {Saillenfest}, {Salaris}, {Salmon}, {Saltas},
  {Samadi}, {Samadi}, {Samra}, {Sanches da Silva}, {Andr{\'e}s S{\'a}nchez
  Carrasco}, {Santerne}, {Santoli}, {Santos}, {Sanz Mesa}, {Sarro},
  {Scandariato}, {Sch{\"a}fer}, {Schlafly}, {Schmider}, {Schneider}, {Schou},
  {Schunker}, {J{\"o}rg Schwarzkopf}, {Serenelli}, {Seynaeve}, {Shan},
  {Shapiro}, {Shipman}, {Sicilia}, {Sierra Sanmartin}, {Sigot}, {Silliman},
  {Silvotti}, {Simon}, {Simoyama Napoli}, {Skarka}, {Smalley}, {Smiljanic},
  {Smit}, {Smith}, {Smith}, {Snellen}, {S{\'o}dor}, {Sohl}, {Solanki},
  {Sortino}, {Sousa}, {Southworth}, {Souto}, {Sozzetti}, {Stamatellos},
  {Stassun}, {Steller}, {Stello}, {Stelzer}, {Stiebeler}, {Stokholm},
  {Storelvmo}, {Strassmeier}, {Str{\o}m}, {Strugarek}, {Sulis}, {{\v{S}}vanda},
  {Szabados}, {Szab{\'o}}, {Szab{\'o}}, {Szuszkiewicz}, {Talens}, {Teti},
  {Theisen}, {Th{\'e}venin}, {Thoul}, {Tiphene}, {Titz-Weider}, {Tkachenko},
  {Tomecki}, {Tonfat}, {Tosi}, {Trampedach}, {Traven}, {Triaud}, {Tr{\o}nnes},
  {Tsantaki}, {Tschentscher}, {Turin}, {Tvaruzka}, {Ulmer}, {Ulmer-Moll},
  {Ulusoy}, {Umbriaco}, {Valencia}, {Valentini}, {Valio}, {Valverde Guijarro},
  {Van Eylen}, {Van Grootel}, {van Kempen}, {Van Reeth}, {Van Zelst},
  {Vandenbussche}, {Vasiliou}, {Vasilyev}, {Vaz de Mascarenhas}, {Vazan}, {Vela
  Nunez}, {Nunes Velloso}, {Ventura}, {Ventura}, {Venturini}, {Trallero},
  {Veras}, {Verdugo}, {Verma}, {Vibert}, {Vicanek Martinez}, {Vida}, {Vigan},
  {Villacorta}, {Villaver}, {Villaverde Aparicio}, {Viotto}, {Vorobyov},
  {Vorontsov}, {Wagner}, {Walloschek}, {Walton}, {Walton}, {Wang}, {Waters},
  {Watson}, {Wedemeyer}, {Weeks}, {Weingril}, {Weiss}, {Wendler}, {West},
  {Westerdorff}, {Westphal}, {Wheatley}, {White}, {Whittaker}, {Wickhusen},
  {Wilson}, {Windsor}, {Winter}, {Lykke Winther}, {Winton}, {Witteck},
  {Witzke}, {Woitke}, {Wolter}, {Wuchterl}, {Wyatt}, {Yang}, {Yu}, {Zanmar
  Sanchez}, {Rosa Zapatero Osorio}, {Zechmeister}, {Zhou}, {Ziemke}, \&
  {Zwintz}}]{Rauer2024}
{Rauer}, H., {Aerts}, C., {Cabrera}, J., {et~al.} 2024, arXiv e-prints,
  arXiv:2406.05447

\bibitem[{{Rehm} {et~al.}(2024){Rehm}, {Mombarg}, {Aerts}, {Michielsen},
  {Burssens}, \& {Townsend}}]{Rehm2024}
{Rehm}, R., {Mombarg}, J. S.~G., {Aerts}, C., {et~al.} 2024, \aap, 687, A175

\bibitem[{{Rendle} {et~al.}(2019){Rendle}, {Buldgen}, {Miglio}, {Reese},
  {Noels}, {Davies}, {Campante}, {Chaplin}, {Lund}, {Kuszlewicz}, {Scott},
  {Scuflaire}, {Ball}, {Smetana}, \& {Nsamba}}]{Rendle2019}
{Rendle}, B.~M., {Buldgen}, G., {Miglio}, A., {et~al.} 2019, \mnras, 484, 771

\bibitem[{{Richer} {et~al.}(2000){Richer}, {Michaud}, \&
  {Turcotte}}]{Richer2000}
{Richer}, J., {Michaud}, G., \& {Turcotte}, S. 2000, \apj, 529, 338

\bibitem[{{Ricker}(2016)}]{TESS}
{Ricker}, G.~R. 2016, in AGU Fall Meeting Abstracts, P13C--01

\bibitem[{{Rogers} \& {Nayfonov}(2002)}]{Rogers2002}
{Rogers}, F.~J. \& {Nayfonov}, A. 2002, \apj, 576, 1064

\bibitem[{{Salaris} \& {Weiss}(2001)}]{Salaris2001}
{Salaris}, M. \& {Weiss}, A. 2001, \aap, 376, 955

\bibitem[{{Seaton}(1997)}]{Seaton1997}
{Seaton}, M.~J. 1997, \mnras, 289, 700

\bibitem[{{Seaton}(2005)}]{Seaton2005}
{Seaton}, M.~J. 2005, \mnras, 362, L1

\bibitem[{{Th{\'e}ado} {et~al.}(2009){Th{\'e}ado}, {Vauclair}, {Alecian}, \&
  {LeBlanc}}]{Theado2009}
{Th{\'e}ado}, S., {Vauclair}, S., {Alecian}, G., \& {LeBlanc}, F. 2009, \apj,
  704, 1262

\bibitem[{{Tinetti} {et~al.}(2018){Tinetti}, {Drossart}, {Eccleston},
  {Hartogh}, {Heske}, {Leconte}, {Micela}, {Ollivier}, {Pilbratt}, {Puig},
  {Turrini}, {Vandenbussche}, {Wolkenberg}, {Beaulieu}, {Buchave}, {Ferus},
  {Griffin}, {Guedel}, {Justtanont}, {Lagage}, {Machado}, {Malaguti}, {Min},
  {N{\'o}rgaard-Nielsen}, {Rataj}, {Ray}, {Ribas}, {Swain}, {Szabo}, {Werner},
  {Barstow}, {Burleigh}, {Cho}, {du Foresto}, {Coustenis}, {Decin}, {Encrenaz},
  {Galand }, {Gillon}, {Helled}, {Morales}, {Mu{\~n}oz}, {Moneti}, {Pagano},
  {Pascale}, {Piccioni}, {Pinfield}, {Sarkar}, {Selsis}, {Tennyson}, {Triaud},
  {Venot}, {Waldmann}, {Waltham}, {Wright}, {Amiaux}, {Augu{\`e}res},
  {Berth{\'e}}, {Bezawada}, {Bishop}, {Bowles}, {Coffey}, {Colom{\'e}},
  {Crook}, {Crouzet}, {Da Peppo}, {Sanz}, {Focardi}, {Frericks}, {Hunt},
  {Kohley}, {Middleton}, {Morgante}, {Ottensamer}, {Pace}, {Pearson},
  {Stamper}, {Symonds}, {Rengel}, {Renotte}, {Ade}, {Affer}, {Alard}, {Allard},
  {Altieri}, {Andr{\'e}}, {Arena}, {Argyriou}, {Aylward}, {Baccani}, {Bakos},
  {Banaszkiewicz}, {Barlow}, {Batista}, {Bellucci}, {Benatti}, {Bernardi},
  {B{\'e}zard}, {Blecka}, {Bolmont}, {Bonfond}, {Bonito}, {Bonomo}, {Brucato},
  {Brun}, {Bryson}, {Bujwan}, {Casewell}, {Charnay}, {Pestellini}, {Chen},
  {Ciaravella}, {Claudi}, {Cl{\'e}dassou}, {Damasso}, {Damiano}, {Danielski},
  {Deroo}, {Di Giorgio}, {Dominik}, {Doublier}, {Doyle}, {Doyon}, {Drummond},
  {Duong}, {Eales}, {Edwards}, {Farina}, {Flaccomio}, {Fletcher}, {Forget},
  {Fossey}, {Fr{\"a}nz}, {Fujii}, {Garc{\'i}a-Piquer}, {Gear}, {Geoffray},
  {G{\'e}rard}, {Gesa}, {Gomez}, {Graczyk}, {Griffith}, {Grodent}, {Guarcello},
  {Gustin}, {Hamano}, {Hargrave}, {Hello}, {Heng}, {Herrero}, {Hornstrup},
  {Hubert}, {Ida}, {Ikoma}, {Iro}, {Irwin}, {Jarchow}, {Jaubert}, {Jones},
  {Julien}, {Kameda}, {Kerschbaum}, {Kervella}, {Koskinen}, {Krijger}, {Krupp},
  {Lafarga}, {Landini}, {Lellouch}, {Leto}, {Luntzer}, {Rank-L{\"u}ftinger},
  {Maggio}, {Maldonado}, {Maillard}, {Mall}, {Marquette}, {Mathis}, {Maxted},
  {Matsuo}, {Medvedev}, {Miguel}, {Minier}, {Morello}, {Mura}, {Narita},
  {Nascimbeni}, {Nguyen Tong}, {Noce}, {Oliva}, {Palle}, {Palmer}, {Pancrazzi},
  {Papageorgiou}, {Parmentier}, {Perger}, {Petralia}, {Pezzuto},
  {Pierrehumbert}, {Pillitteri}, {Piotto}, {Pisano}, {Prisinzano}, {Radioti},
  {R{\'e}ess}, {Rezac}, {Rocchetto}, {Rosich}, {Sanna}, {Santerne}, {Savini},
  {Scandariato}, {Sicardy}, {Sierra}, {Sindoni}, {Skup}, {Snellen}, {Sobiecki},
  {Soret}, {Sozzetti}, {Stiepen}, {Strugarek}, {Taylor}, {Taylor}, {Terenzi},
  {Tessenyi}, {Tsiaras}, {Tucker}, {Valencia}, {Vasisht}, {Vazan}, {Vilardell},
  {Vinatier}, {Viti}, {Waters}, {Wawer}, {Wawrzaszek}, {Whitworth}, {Yung},
  {Yurchenko}, {Osorio}, {Zellem}, {Zingales}, \& {Zwart}}]{Tinetti2018}
{Tinetti}, G., {Drossart}, P., {Eccleston}, P., {et~al.} 2018, Experimental
  Astronomy, 46, 135

\bibitem[{{Tinetti} {et~al.}(2022){Tinetti}, {Eccleston}, {Lueftinger},
  {Salvignol}, {Fahmy}, \& {Alves de Oliveira}}]{tinetti2021}
{Tinetti}, G., {Eccleston}, P., {Lueftinger}, T., {et~al.} 2022, in European
  Planetary Science Congress, EPSC2022--1114

\bibitem[{{Valle} {et~al.}(2014){Valle}, {Dell'Omodarme}, {Prada Moroni}, \&
  {Degl'Innocenti}}]{Valle2014}
{Valle}, G., {Dell'Omodarme}, M., {Prada Moroni}, P.~G., \& {Degl'Innocenti},
  S. 2014, \aap, 561, A125

\bibitem[{{Valle} {et~al.}(2015){Valle}, {Dell'Omodarme}, {Prada Moroni}, \&
  {Degl'Innocenti}}]{Valle2015}
{Valle}, G., {Dell'Omodarme}, M., {Prada Moroni}, P.~G., \& {Degl'Innocenti},
  S. 2015, \aap, 575, A12

\bibitem[{{Verma} \& {Silva Aguirre}(2019)}]{Verma2019}
{Verma}, K. \& {Silva Aguirre}, V. 2019, \mnras, 489, 1850

\end{thebibliography}


\begin{appendix}

\onecolumn
\section{Implementing the SVP method in MESA}
\label{ap:how_SVP_implementation}

\begin{figure*}[h]
    \centering
    \includegraphics[width=0.85\linewidth]{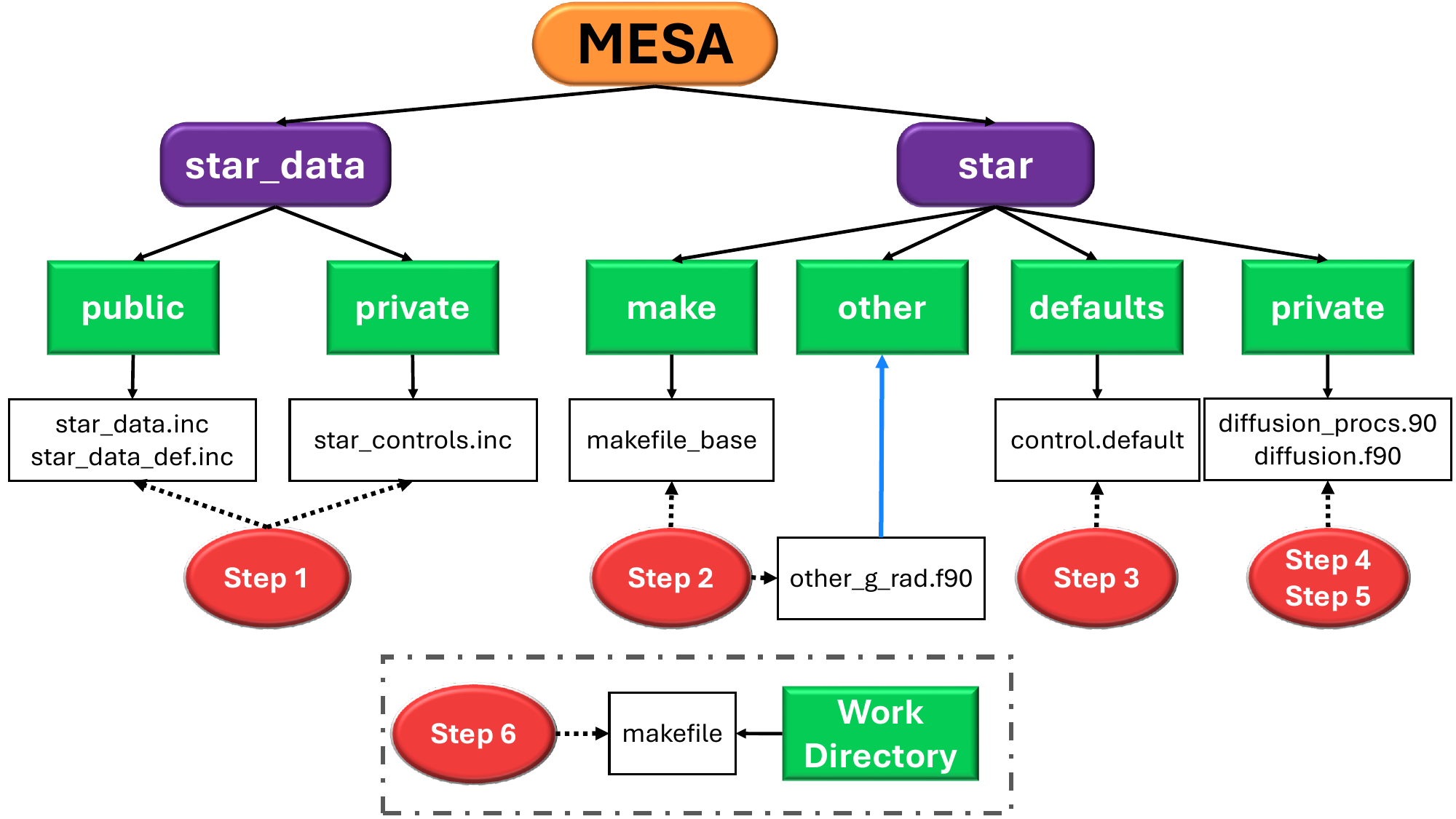}
    
    \caption{Schematic of the MESA modules and folders that need to be modified to implement SVP.}
    \label{fig:SVP_imp_sch}
\end{figure*}

The SVP method was developed by G. Alecian and F. LeBlanc \citep{LeBlanc2004,alecian20} and the routines to compute radiative accelerations are publicly available\footnote{\url{https://gradsvp.obspm.fr/index.html}}. These routines require only a few input parameters to compute radiative accelerations, such as $T$, $\rho$, $P$, $L$, $r$, $m$, and $X_\mathrm{i}$, the mass fraction of the elements being followed. The method is currently only applicable to main sequence stars (as the tables are only prepared for this stage). The implementation in MESA is not as simple as for turbulent mixing, since the code does not provide an external hook for the radiative accelerations. In this case, it is necessary to modify the MESA source code directly. To minimize code changes, we added a new hook into MESA to be used as an alternative to the default radiative accelerations. For the implementation, we took the following steps (a schematic is presented in Fig.~\ref{fig:SVP_imp_sch}):

\begin{enumerate}
    \item Add a new hook that creates null functions in the files \textit{star\_data.inc}, \textit{star\_data\_def.inc}, and \textit{star\_controls.inc} in the \textit{star\_data} folder;
    \item Create a new hook file/module called \textit{other\_g\_rad.f90} in the folder \textit{other} and add it in the MESA \textit{makefile\_base} for compilation in the \textit{make} folder. 
    All in the \textit{star} folder of MESA;
    \item Add in the \textit{control.default} file the routine and set it off by default in MESA;
    \item In the radiative acceleration file (in \textit{diffusion\_procs.f90}), add a condition to select whether to use the hook or the default routine;
    \item Add a condition in the diffusion routine that allows us to turn the computation of the Rosseland mean opacity with monochromatic opacity on or off when using SVP (in \textit{diffusion.f90} and \textit{diffusion\_procs.f90}). MESA by default does not allow the use of radiative accelerations if the monochromatic opacities are not used;
    \item Add a folder that contains the SVP routine, and add its directory in the \textit{makefile} for the models;
    \item Finally, reinstall MESA after making the changes.
\end{enumerate}
This gives access to an ``outside'' hook to implement all the different functions needed to call the SVP routine, without major changes to the source code of MESA. The routine we implemented are publicly available\footnote{\url
{https://doi.org/10.5281/zenodo.14678036}}.

\FloatBarrier
\clearpage
\section{Absolute differences}
\label{ap:abs_results}

\begin{figure}[h]
    \centering
    \includegraphics[width=1\columnwidth]{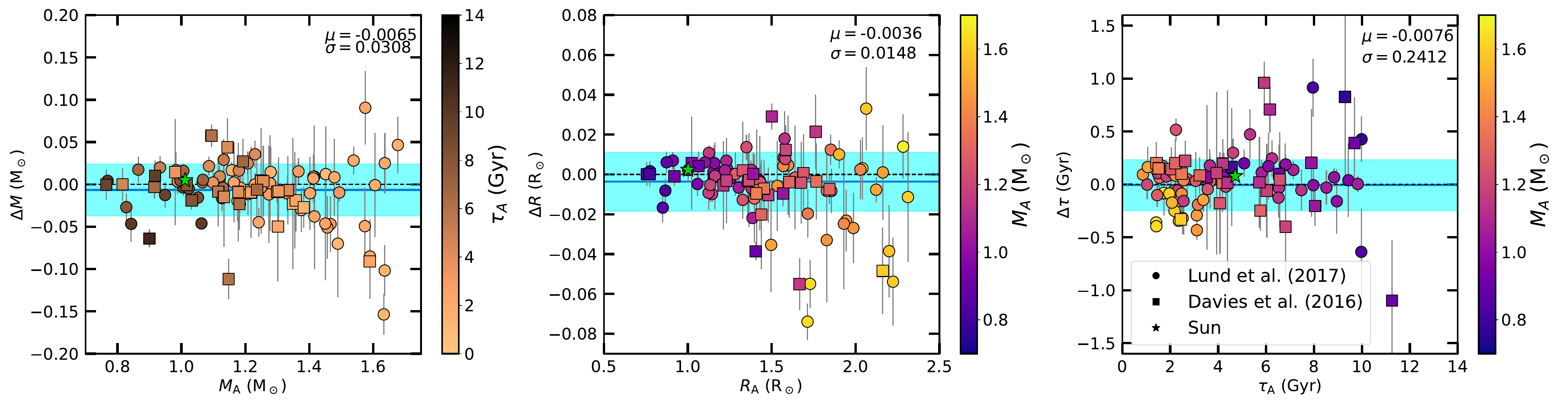}
    \caption{Absolute difference for mass (left panel), radius (middle panel), and age (right panel) between grids A and C. The solid blue line indicates the bias, and the blue region is the 1$\sigma$ of the standard deviation. Each point is color-coded with the corresponding reference age (left panel) and mass (middle and right panels).}
    \label{fig:stellar_prop_SVP_nodiff_abs}
\end{figure}

\begin{figure}[h]
    \centering
    \includegraphics[width=1\columnwidth]{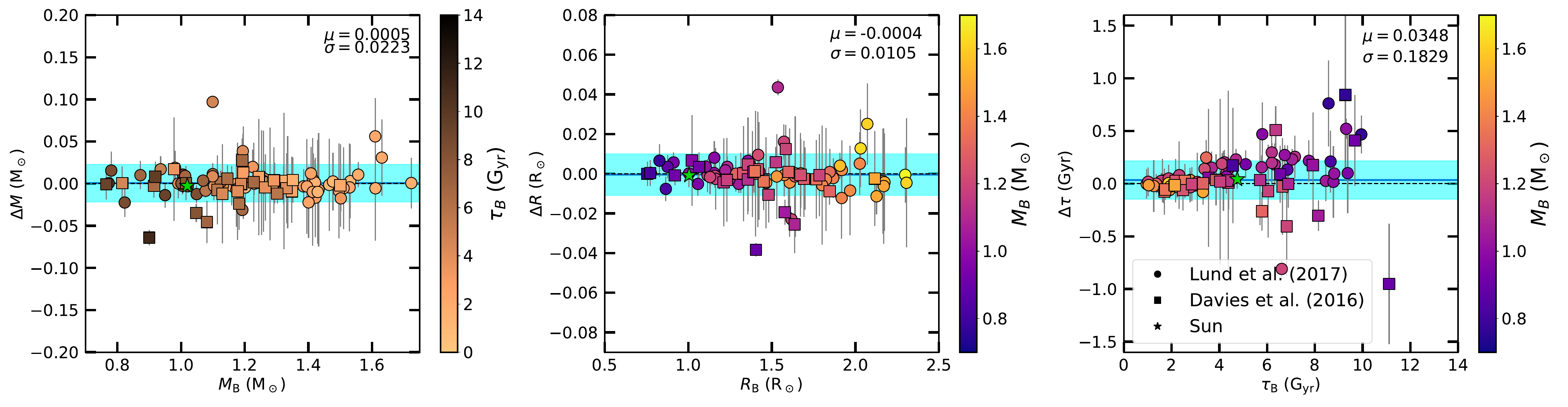}
    \caption{Same as Fig.\ B.1 but for the differences between grids B and C. }
    \label{fig:stellar_prop_SVP_Dturb_abs}
\end{figure}

Figures \ref{fig:stellar_prop_SVP_nodiff_abs} and \ref{fig:stellar_prop_SVP_Dturb_abs} show the absolute difference of the derived values. Figure \ref{fig:stellar_prop_SVP_nodiff_abs} are the results between grids C and A, and Fig.~\ref{fig:stellar_prop_SVP_Dturb_abs} the results between grids C and B.

\end{appendix}


\end{document}